\pdfoutput=1

\documentclass[prb,aps,twocolumn,amsmath,amssymb,floatfix,
superscriptaddress]{revtex4}
\usepackage{graphics,epstopdf}
\usepackage{epsf,graphics,graphicx}
\usepackage{color}
\usepackage{soul}
\usepackage{mathrsfs}

\usepackage{comment}
\usepackage{subcaption}
\usepackage[colorlinks=true,linkcolor=blue,citecolor=blue,urlcolor=blue]{hyperref}

\begin{document}

\title{Transport signatures of topological commensurate off-diagonal Aubry-André-Harper Chain }

\author{Arpita Koley}

\email{2708.arpita@gmail.com}

\affiliation{School of Physical Sciences, Indian Association for the Cultivation of Science, 2A and 2B Raja S.C. Mullick Road, Jadavpur, Kolkata 700032, India}

\begin{abstract}
We study the interplay between quantum transport and topology in a one-dimensional off-diagonal commensurate Aubry-André-Harper (AAH) chain. The model, formulated within AAH framework, effectively represents a one-dimensional lattice with two competing commensurate modulations, supporting two distinct types of topological edge modes: zero-energy states in the central gapless region and quantum Hall (QH) edge states bridging the gapped bulk bands. These edge modes govern the transport behavior and give rise to sharp variations in transmission across the corresponding gap-closing transitions. A pronounced even-odd effect further emerges, where chains with an odd number of sites exhibit nearly perfect zero-energy transmission at the Dirac points, independent of system-lead coupling, system size, or modulation strength; a robust signature of ballistic transport. To capture the influence of environmental decoherence, we also incorporate Büttiker dephasing probes, which enable a phenomenological description of inelastic scattering and reveal how dephasing modifies, and in some regimes enhances, coherent transport.

\end{abstract}

\maketitle

\section{Introduction}
Quasiperiodic lattice models have long served as paradigmatic systems for understanding the interplay between order, disorder, and topology in low-dimensional quantum systems.~\cite{Mob,AA,Har,Multif,scal,SDS2,Qua} Among them, the AAH model occupies a central position, providing a minimal yet rich framework to study electron localization at finite quasiperiodic disorder strength with energy dependent or independent mobility edge~\cite{Mob,Mob3,SDS2,exp1,exp2,abr5}, and quantum transport in one dimension. Originally introduced to describe electrons in a one-dimensional lattice under a modulated potential, the model also captures the physics of a two-dimensional electron gas on a square lattice subject to a uniform magnetic flux through the celebrated Harper–Hofstadter correspondence~\cite{Har,Hofs,2ds,QH}.\vspace{0.15cm}

Sharp metal–insulator transition at a critical modulation strength is a signature of the  incommensurate limit of the AAH potential imposed on a periodic lattice.~\cite{MiT} The eigenstates evolve from extended Bloch wavefunctions to exponentially localized states, with the critical point characterized by multifractal behavior.~\cite{Multif} Transport in this regime has been extensively studied~\cite{MiT,exp3}, revealing ballistic conduction for extended states, exponentially suppressed transport in the localized regime, and anomalous, sub-diffusive scaling of transmission at criticality.~\cite{trans} Furthermore, adiabatic modulation of the AAH Hamiltonian can induce quantized charge pumping through QH edge states, known as Thouless pumping.~\cite{Thou2,Zil1,Zil2} This phenomena directly connects the topological Chern number\cite{Cn} of the two-dimensional ancestor model to measurable transport in one dimension.\vspace{0.15cm}

When the modulation becomes commensurate, i.e., the modulation parameter $b=p/q$ is rational, the system hosts q sub-bands separated by topological energy gaps, connected by localized edge states. In contrast to the incommensurate case\cite{Multif,scal,SDS2}, the commensurate AAH chain~\cite{SDS1} can exhibit non-trivial topology with zero-energy Majorana modes, depending on whether the modulation is applied to the onsite potential (diagonal) or the hopping amplitudes (off-diagonal). The off-diagonal commensurate AAH model, in particular, serves as a natural generalization of the Su-Schrieffer-Heeger (SSH) model\cite{SSH}, supporting both topologically protected zero-energy and QH edge modes.~\cite{MB1,Wind}\vspace{0.15cm}

For specific rational values of the modulation parameter, such as $b=1/2$ and $1/4$, the energy spectrum exhibits qualitatively distinct topological features\cite{MB1,SDS1,Topo}. The case $b=1/2$ corresponds to a gapless phase hosting zero-energy edge states, whereas $b=1/4$ exhibits a single topological transition characterized by the closing and reopening of the central energy gap separating the bulk bands.~\cite{SDS1} In contrast, for 
$b=1/6$, the spectrum reveals multiple gap-closing and reopening events, indicating a richer hierarchy of topological transitions. The first of these transitions reproduces the behavior observed in the $b=1/4$ case, while subsequent transitions involve adjacent bulk bands, leading to the reappearance of QH edge states reminiscent of the two-dimensional Chern insulator. On the other hand, higher-order commensurations ($b=1/8$, $b=1/10$) produce additional, more intricate sequences of band inversions without introducing new types of localized edge states. Thus, the commensurate off-diagonal AAH chain with $b=1/6$ represents the minimal configuration capable of capturing both zero-energy and QH edge physics within a single one-dimensional framework.\vspace{0.15cm}

From a transport perspective, such rich spectral and topological structures yield striking physical consequences. In contrast to the incommensurate case, the commensurate off-diagonal AAH chain hosts fully extended Bloch bands across the entire spectrum. However, electronic transport through these bulk bands remains highly sensitive to the presence and nature of the edge states. Furthermore, finite-size geometry and boundary conditions play a decisive role: chains with an odd number of sites exhibit nearly perfect zero-energy transmission within specific parameter regimes, while even-site systems display suppressed conductance governed by the terminating bond characteristics.\vspace{0.15cm}

In this work, we investigate the detailed connection between transport and topology in the commensurate off-diagonal AAH model with rational value $b=1/6$. By analyzing the evolution of transmission spectra across topological transitions, we uncover distinct signatures in bulk band conduction along with even-odd parity effects.~\cite{SDS1,Topo} Furthermore, to account for realistic environmental effects, we introduce Büttiker dephasing~\cite{MBP1,MBP2,MBP3,BP1,BP2,BP3,BP0} probes to model inelastic scattering and decoherence. Electrons loose phase coherence upon entering fictitious voltage probes and re-enter the conductor while maintaining zero net current flow. This inclusion allows us to explore environment-assisted transport, where decoherence can enhance conduction and become comparable to the coherent limit. \vspace{0.15cm}

Overall, our study establishes the commensurate off-diagonal AAH chain as a versatile testbed to explore the combined effects of topology, finite-size geometry, and dephasing in one-dimensional quantum transport. The insights obtained here are relevant for engineered quantum systems, including ultracold atom lattices, photonic waveguides, and molecular electronic devices, where controlled quasiperiodicity and dephasing can be realized experimentally.~\cite{exp0,exp1,Edge,Thoupump}\vspace{0.15cm}

We organize the paper as follows. In Sec.~\ref{sec:setup}, we describe the detailed junction setup and present the model Hamiltonian of the commensurate off-diagonal AAH chain, including its coupling to external electrodes and Büttiker probes used to simulate environmental dephasing. Section~\ref{sec:numresult} discusses the numerical results, where we analyze the interplay between quantum transport and gap closing transitions, highlighting the roles of bulk and edge states under different commensurate modulations. Finally, Sec.~\ref{sec:remark} summarizes our main findings and provides concluding remarks.


  
\section{Formulation of TB Hamiltonian}
\label{sec:setup}
\subsection{Junction setup and the Hamiltonian}
We consider a one-dimensional conductor (the commensurate off-diagonal AAH chain) connected to two semi-infinite metallic leads and to local infinite impedance Büttiker probes that model dephasing. The total Hamiltonian of the junction, written in tight-binding approximation including only nearest-neighbor hopping, is structurally the same as the model studied by M. Saha \textit{et al.}\cite{BP2}. It looks like\cite{BP1,BP2,BP3}- 
\begin{equation}
H=H_C + H_L + H_{CL} + H_P+H_{CP}
\label{equ1}
\end{equation}
where the subscripts denote: conducting AAH chain (C), leads (L = S (source) and D (drain)), conductor-lead coupling (CL), fictitious probe reservoirs (P) and conductor-probe coupling (CP). Below we specify each term and its physical meaning. The conductor Hamiltonian $H_C$ includes site-dependent off-diagonal (hopping) modulation and it is written in the following form,
\begin{equation}
H_C=\sum_{i=1}^N t_0(1+ \delta_t \cos(2\pi b i + \phi_{\lambda}))c^{\dagger}_{i+1}c_i + h.c.
\label{equ2}
\end{equation}
Here $c_i$ ($c_i^{\dagger}$) is fermionic annihilation (creation) operator on site $i$ of the conductor, $t_0$ is the unmodulated nearest-neighbour hopping, $\delta_t$ is the dimensionless AAH modulation strength, $b=p/q$ represents the rational (commensurate) modulation  and $\phi_{\lambda}$ is the AAH phase. The parameters` $b$' and `$\phi_{\lambda}$' acquire a direct physical meaning through their well-known correspondence to the two-dimensional Hofstadter model\cite{Zil1}. Specifically,  $\pi b$ plays the role of an effective magnetic flux penetrating each plaquette of the Hofstadter lattice, while $\phi_{\lambda}$ maps onto the transverse momentum $k_y$. For the modulation used in this work, $b=1/6$, the equivalent Hofstadter flux is therefore $\phi_{\lambda}=\pi/6$  (including diagonal hopping for this off-diagonal commensurate limit\cite{2ds}). The chain is composed of N sites; the parity of N (even/odd) matters for the zero-energy transmission properties described in the numerical result section.
Each lead is modeled as a semi-infinite, single-channel metallic chain:
\begin{subequations}
\begin{align}
H_S & = \sum_{i\leq 0} a_i^{\dagger} \epsilon_l a_i + \left(a_{i+1}^{\dagger}t_l a_i
+ h.c. \right) \label{equ3a} \\
H_D & = \sum_{i\geq N+1} b_i^{\dagger}\epsilon_l b_i +\left(b_{i+1}^{\dagger}t_l b_i 
+ h.c. \right) \label{equ3b}
\end{align}
\end{subequations}
Operators $a_i$ and $b_i$ act identical to $c_i$ on source (S) and drain (D) sites respectively; and $\epsilon_l$ and $t_l$ are the on-site energy and nearest-neighbour hopping amplitude in the leads. Leads are assumed reflectionless at the CL interface. 
The coupling between conductor and leads is taken at the two ending sites of the central conductor:   
\begin{equation}
H_{CL}= H_{CS}+H_{CD}=c_1^{\dagger}\tau_Sa_{0} + c_N^{\dagger}\tau_D b_{N+1} + h.c.
\label{equ4}
\end{equation} 
where $\tau_{S(D)}$ are the coupling amplitude between the $1$($N$)-th site of biperiodic chain and the first site of source(drain).\vspace{0.15cm}

To account for incoherent processes, we attach $N$  identical local probes to each site of the conductor. Fourth term of the Eq.\ref{equ1} represents the sum of all semi-infinite local voltage probes.  
\begin{equation}
H_P= \sum_{i=1}^N H_P^n =\sum_{i=1}^N \sum_j  d_{i,j}^{\dagger}\epsilon_l d_{i,j} + \left(d_{i,j+1}^{\dagger}t_l d_{i,j} 
+ h.c. \right) 
\label{equ5}
\end{equation}
In fermionic annihilation operator $d_{ij}$, $i$ represents the site number of AAH chain connecting to B\"{u}ttiker probe and $j$ is the site index of $i$-th voltage probe. The coupling between the conductor and the probes is given by
\begin{equation}
H_{CP}=\sum_{i=1}^N H_{CP}^i =\sum_{i=1}^N c_i^{\dagger}\tau_p d_{i1} + h.c.
\label{equ6}
\end{equation}
where $\tau_p$ characterizes the voltage probe-conductor coupling strength and effectively controls the dephasing rate.

\subsection{Theoretical formulation}
To incorporate the influence of all semi-infinite Büttiker probes without explicitly using the full Hilbert space, their effects are embedded through additional self-energy corrections. The effective retarded single particle Green’s function of the reduced system, including the contributions from the source, drain, and dephasing probes, is expressed as \cite{MBP3,BP1,BP2,BP0}-
\begin{equation}
G^r (E)=\left(EI-H_C-\Sigma_S (E)-\Sigma_D (E) - \sum_{i=1}^N \Sigma_P^i (E) \right)^{-1}
\label{equ7}
\end{equation} 
where $E$ denotes the energy of the injected electron and $I$ is the identity matrix of dimension ($N\times N$). $\Sigma_S$, $\Sigma_D$, $\Sigma_P^i$ correspond to the self-energy matrices arising from the couplings between the conductor and the source, drain, and the $i$-th Büttiker probe, respectively. 
Self-energy term can be expressed analytically as\cite{gf1,gf2,gf4,gf5}
\begin{equation}
\Sigma_m (E)= \dfrac{\tau_m^2}{2 t_m^2} \left[E-\epsilon_m-i \sqrt{4t_m^2-(E-\epsilon_m)^2}\right]
\label{equ8}
\end{equation}
,where $m=S,D,P$. $\epsilon_m=\epsilon_l$ and $t_m=t_l$ for all m.  
Since the present analysis is performed at zero temperature $T=0$K, the temperature independent chemical potentials of the source $\mu_S$, drain $\mu_D$, and the local probes $\mu_P^i (i \in \left[1,N\right])$ determine the occupation of electronic states. The probe potentials $\mu_P^i$ are determined self-consistently to enforce charge conservation, such that the net current through each Büttiker probe vanishes. This condition ensures a steady-state current from S to D through the extended Bloch bands of conductor and this phenomena is well studied in literature~\cite{BP1,BP2,BP0}. In this way, the formalism captures partially incoherent transport and provides a simple yet powerful means to study the effective transmission from source to drain, given by,
\begin{equation}
T_{eff} (E)=T_{S,D}(E)+\sum_{i,j=1}^N T_{D,i}(E) W_{i,j}^{-1} T_{j,S}(E)
\label{equ9}
\end{equation}
First term of the effective transmission in Eq.~\ref{equ9} corresponds to the phase-coherent component of transport, whereas the remaining contribution arises from electrons that undergo at least one inelastic scattering event before reaching the drain.
$T_{i,j}$ denotes the transmission probability of an electron propagating from the $i$ to $j$-th electrode.
Elements of $W$ matrix are denoted by- 
\begin{equation}
W_{i,j}=\left[(1-R_{i,i}) \delta_{ij} - T_{i,j} (1-\delta_{ij})\right]
\label{equ10}
\end{equation}
The diagonal element $R_{i,i}= 1-\sum_{j\neq i} T_{i,j}$ represents the total reflection from the $i$-th lead. Using the Green’s function formalism (Eq.\ref{equ7}) together with the Fisher-Lee expression\cite{gf3}, the energy-resolved transmission probability $T_{i,j}(E)$ between the $i$ and $j$-th terminals can be written as, 
\begin{equation}
T_{i,j}(E)=\mbox{Tr}\left[\Gamma_i (E) G^r(E)\Gamma_j (E)(G^r (E))^{\dagger} \right]
\label{equ11}
\end{equation}
where $\Gamma_i (E)$ is the coupling matrix associated with $i$-th probe. This coupling matrix is found from their respective self-energy matrix via the relation,
\begin{equation}
\Gamma_{i} (E)=i\left[
\Sigma_{i} (E)-
\Sigma_{i}^{\dagger} (E) \right]=-2 Im \left[\Sigma_i (E)\right] .
\label{equ12}
\end{equation}
Substituting Eqs.(\ref{equ10})-(\ref{equ12}) into Eq.\ref{equ9}, we obtain the effective transmission incorporating the effect of environmental dephasing through all Büttiker probes.

\section{Numerical results and discussion}
\label{sec:numresult}
In this section, we present a detailed numerical analysis of the transport characteristics associated with the topologically non-trivial gapless regime of the commensurate off-diagonal AAH chain, focusing primarily on the case $b=1/6$. The analysis is organized into four parts addressing: (a) the influence of the off-diagonal modulation strength $\delta_t$ and analytical estimation of the first transition point, (b) the role of the modulation phase $\phi_{\lambda}$, (c) even-odd lattice-size effects on quantum transport, and  (d) the impact of environmental decoherence introduced through Büttiker dephasing probes. Except for subsection (d), all results correspond to coherent transport ($\tau_p=0$ eV) i.e., in the absence of dephasing. 
Unless otherwise stated, the system parameters are fixed as follows: the coupling between the leads and the conductor is set to
$\tau_s=\tau_d=0.75$ eV, the hopping amplitude within the semi-infinite leads is $t_l=2.5$ eV, and the nearest-neighbor hopping in conductor without AAH potential is $t_0=1.0$ eV. Variations from these default parameters are explicitly mentioned in the respective discussions.
\begin{figure}[ht]
    \centering
    \begin{subfigure}[t]{0.23\textwidth}
        \includegraphics[width=\textwidth]{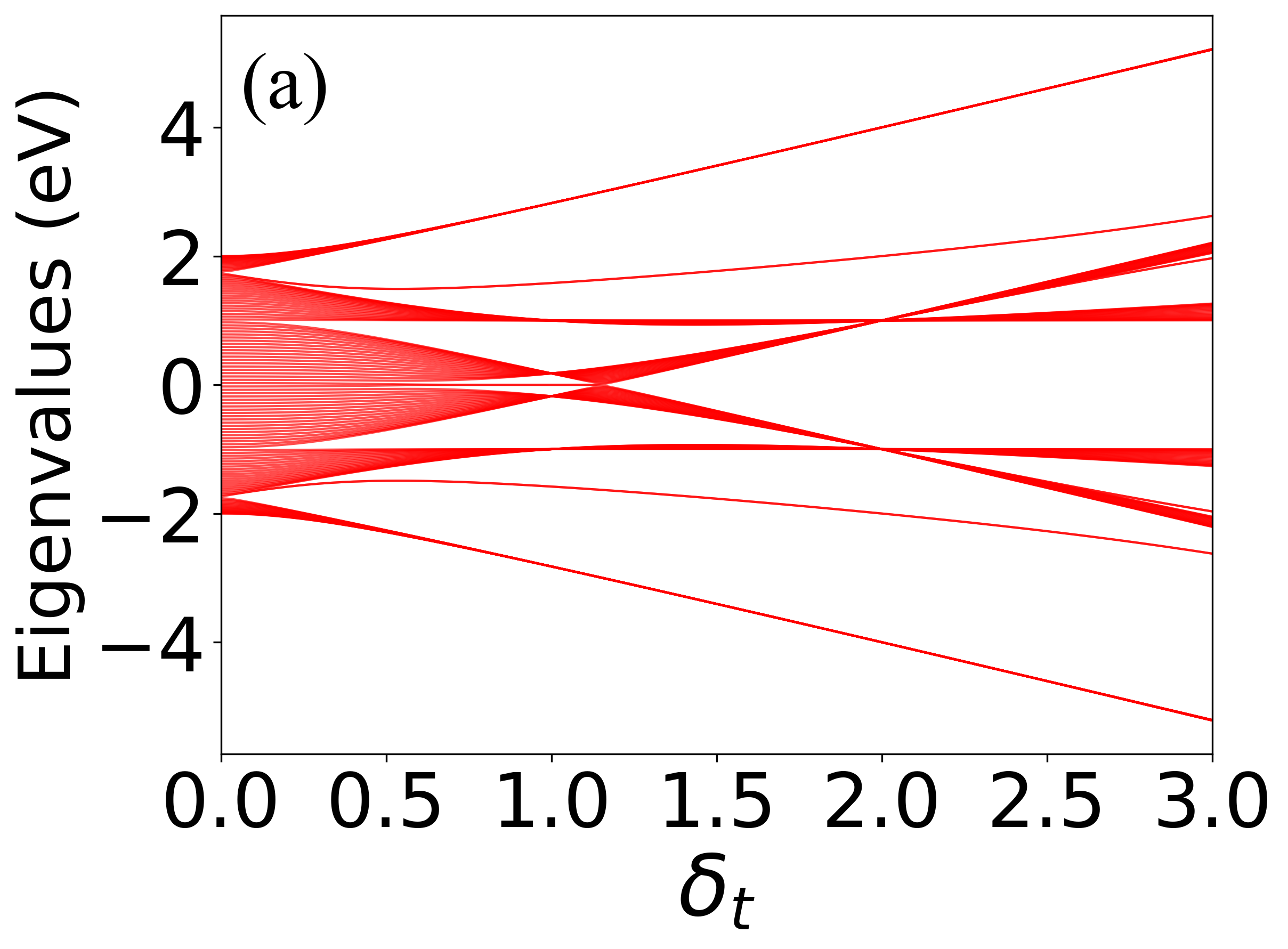}
    \end{subfigure}
    \begin{subfigure}[t]{0.23\textwidth}
        \includegraphics[width=\textwidth]{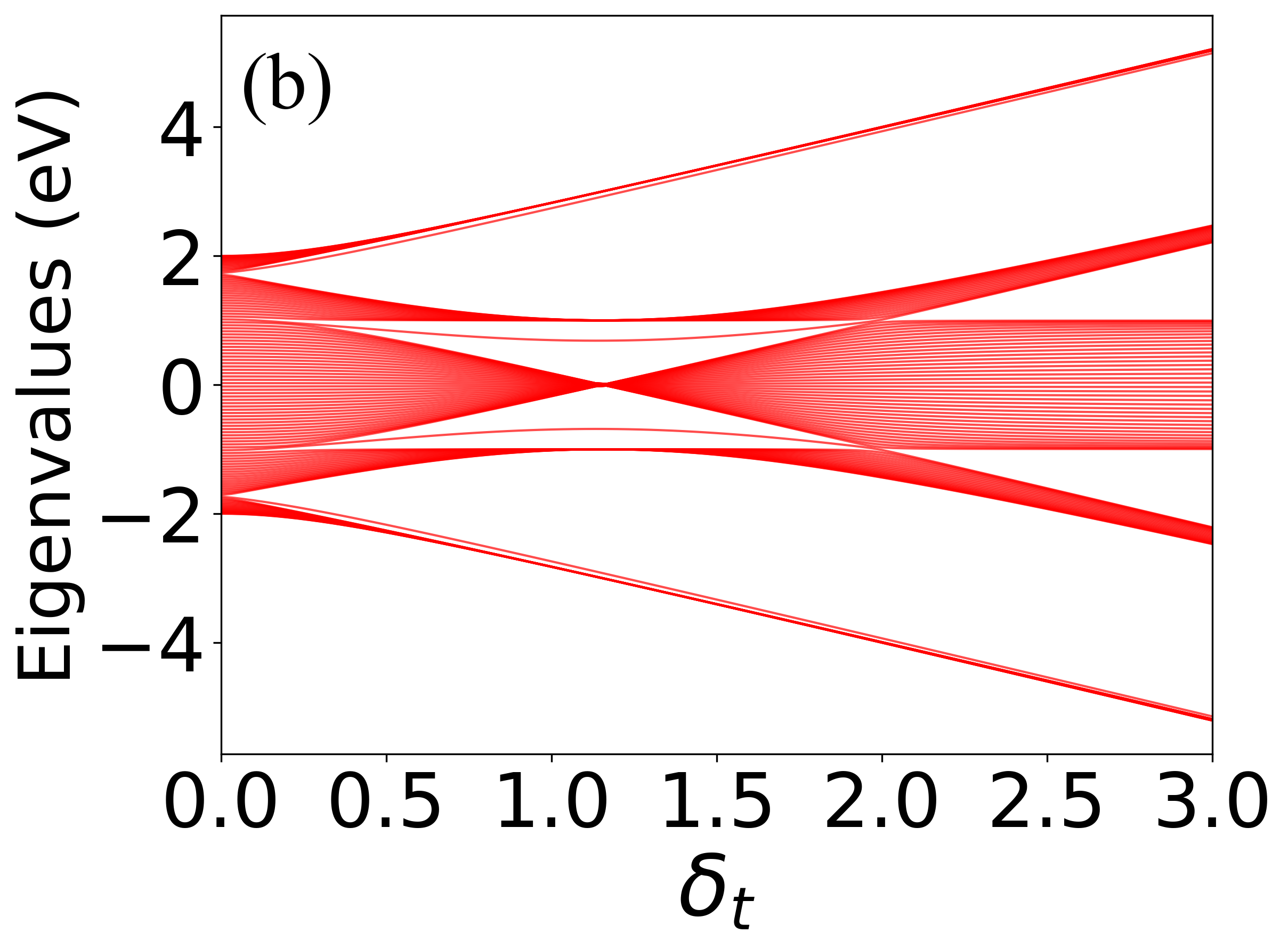}
    \end{subfigure}
\caption{Energy spectra of the isolated off-diagonal commensurate AAH chain with modulation strength $\delta_t$ at $b=\frac{1}{6}$ and lattice size $N=120$ for two representative modulation phases: $\phi_{\lambda}=0$ in (a) and $\pi/2$ in (b). Zero-energy edge states merge with central bulk bands in (a), while a complete band narrowing occurs at $E=0$ eV in (b) at first transition point $\delta_t=W_{c1}=\frac{2}{\sqrt{3}}$. The closing and reopening of the central bulk gap are observed at $\delta_t=W_{c2}=2$ in both configurations.}
    \label{fig1}
\end{figure}

\subsection{Effect of off-diagonal modulation strength $\delta_t$}
\begin{figure}[ht]
    \centering
    \begin{subfigure}[t]{0.2\textwidth}
        \includegraphics[width=\textwidth]{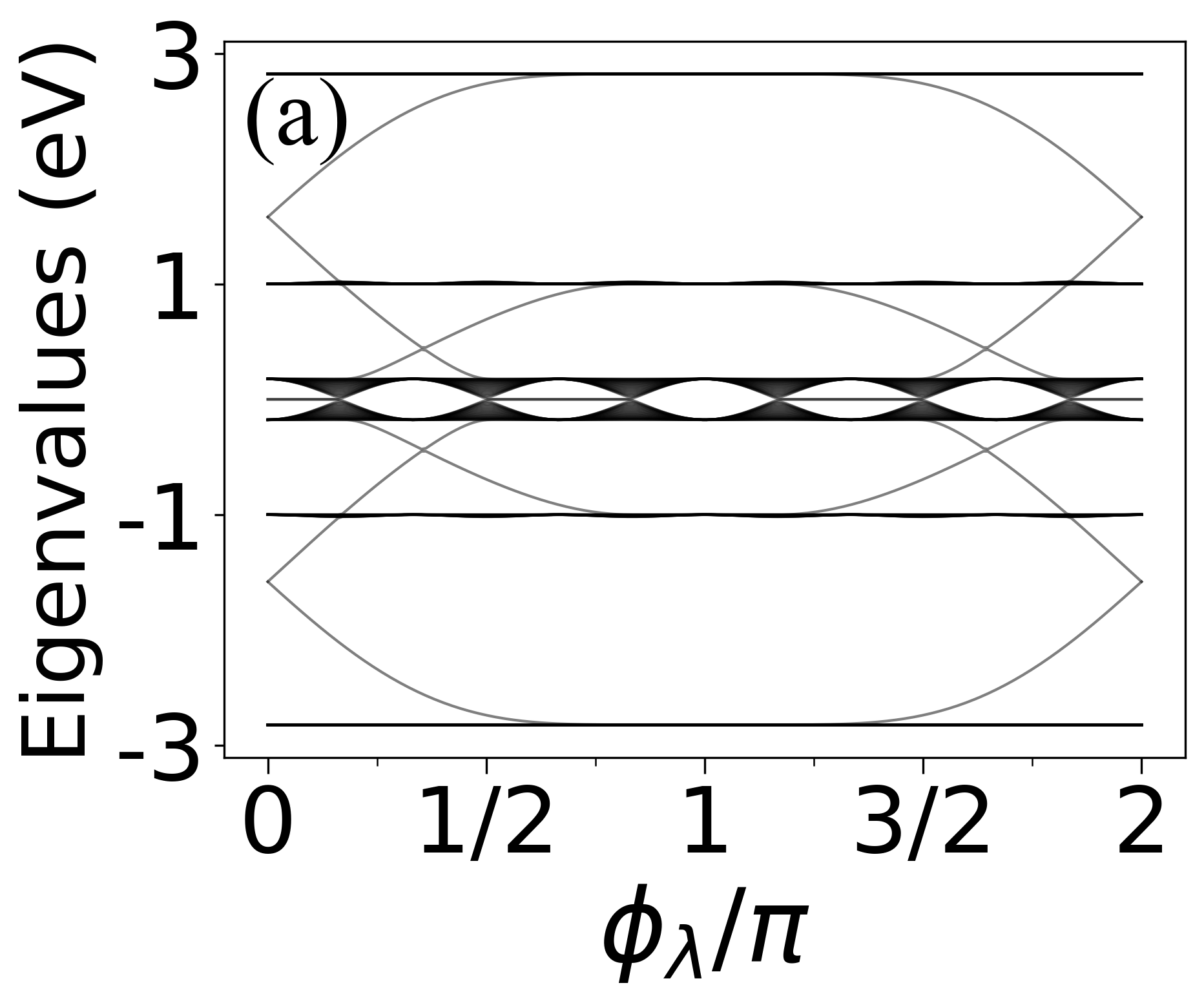}
    \end{subfigure}
    \begin{subfigure}[t]{0.2\textwidth}
        \includegraphics[width=\textwidth]{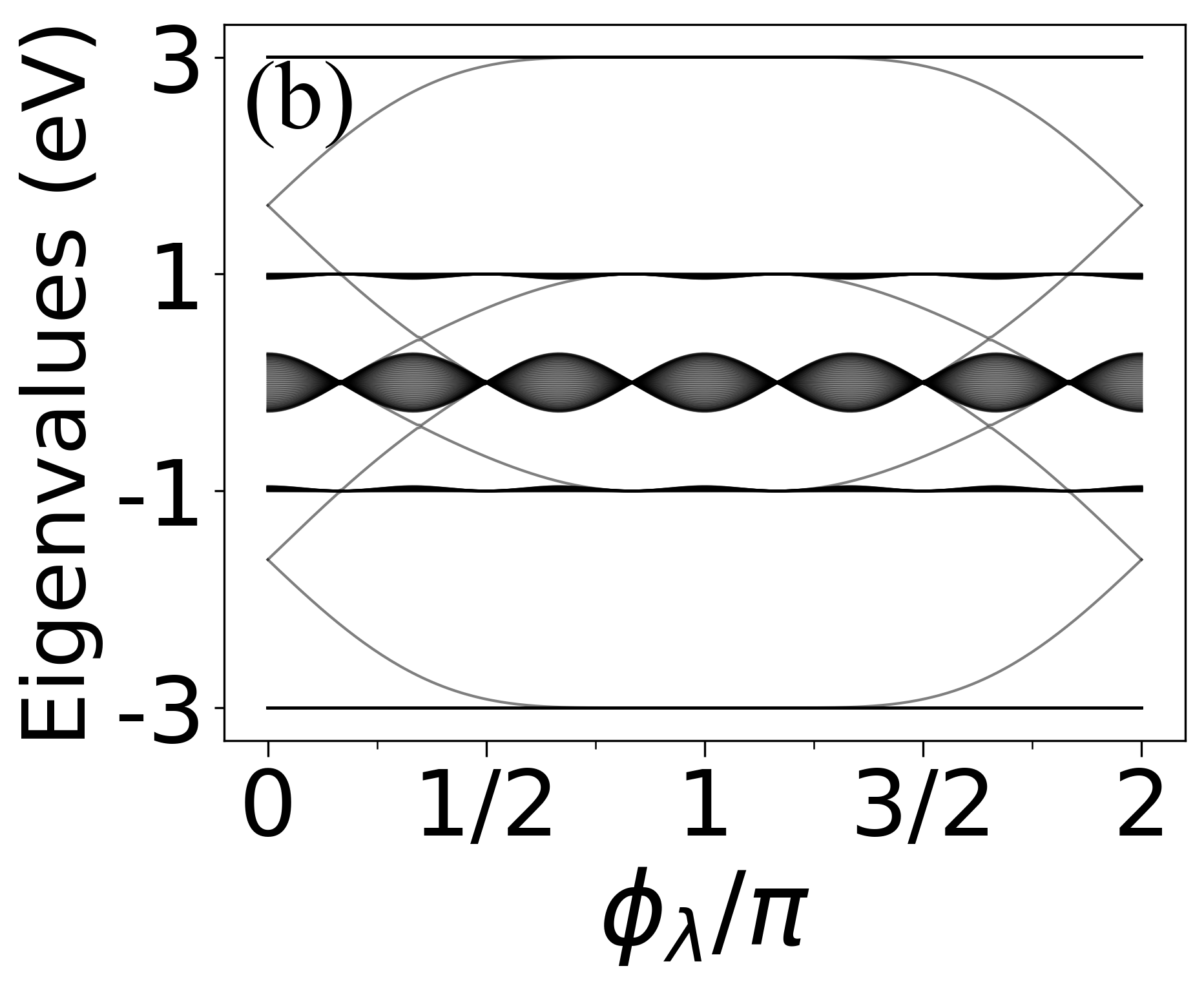}
    \end{subfigure}
    \begin{subfigure}[t]{0.2\textwidth}
        \includegraphics[width=\textwidth]{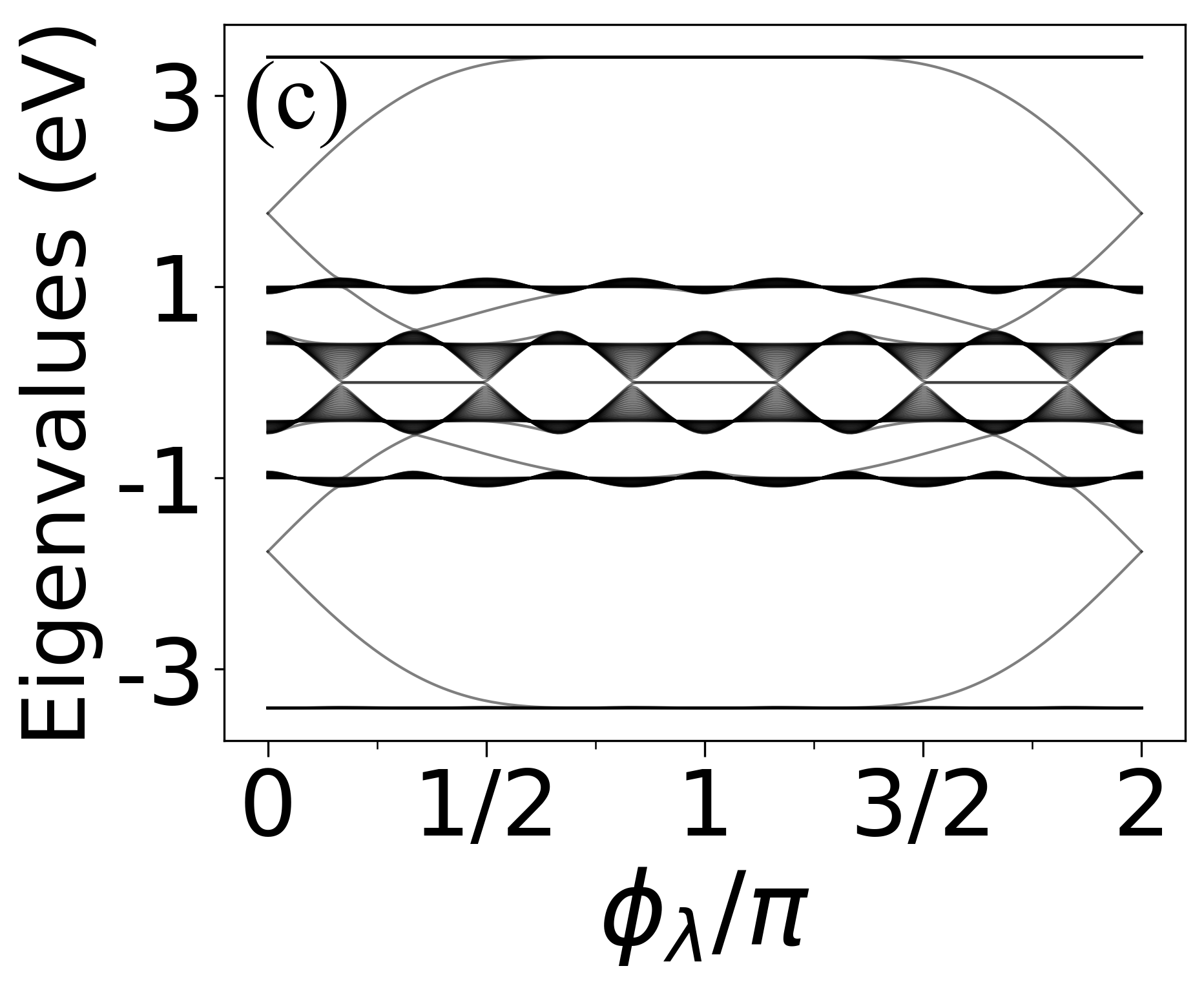}
    \end{subfigure}
    \begin{subfigure}[t]{0.2\textwidth}
        \includegraphics[width=\textwidth]{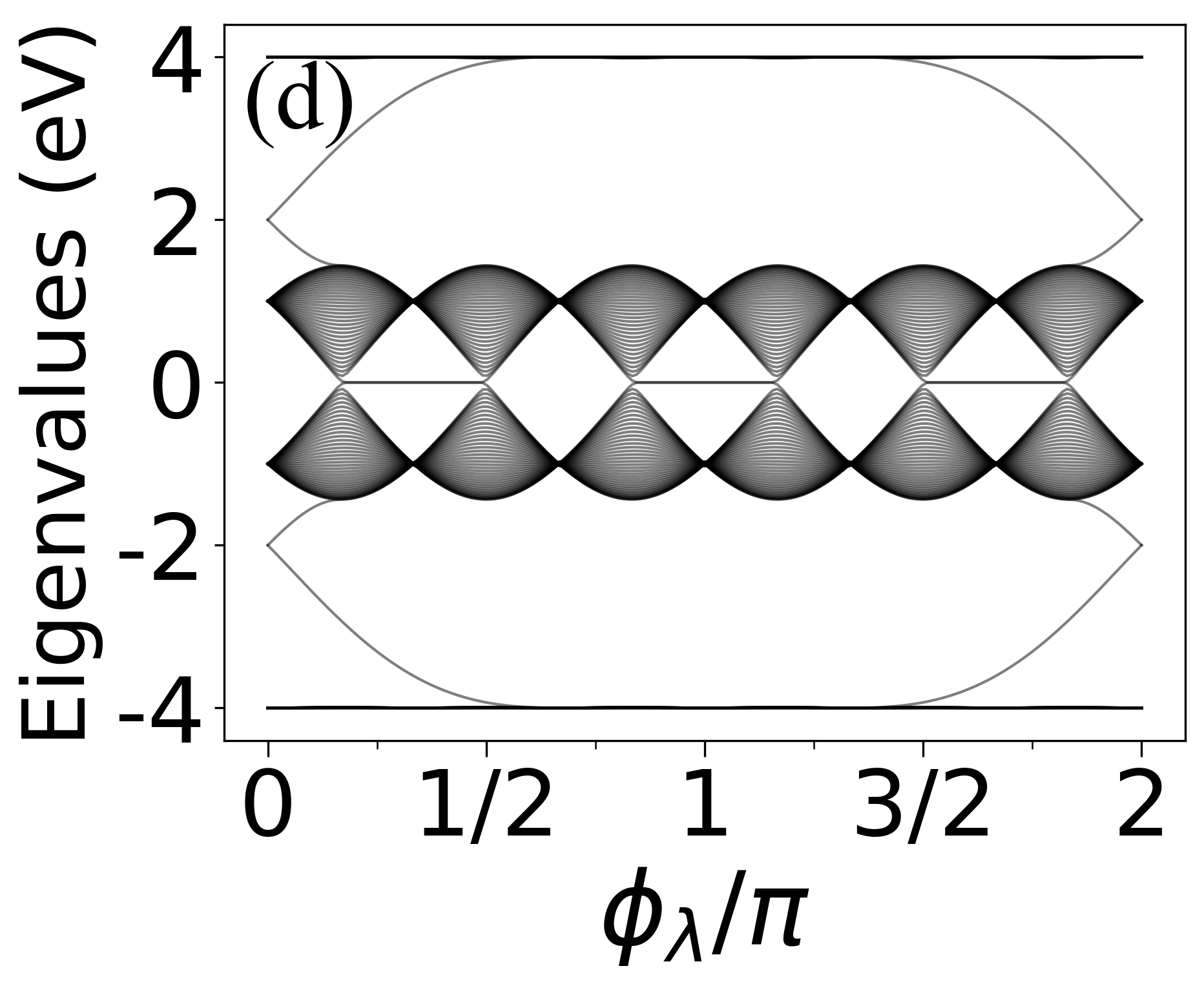}
    \end{subfigure}
    \begin{subfigure}[t]{0.2\textwidth}
        \includegraphics[width=\textwidth]{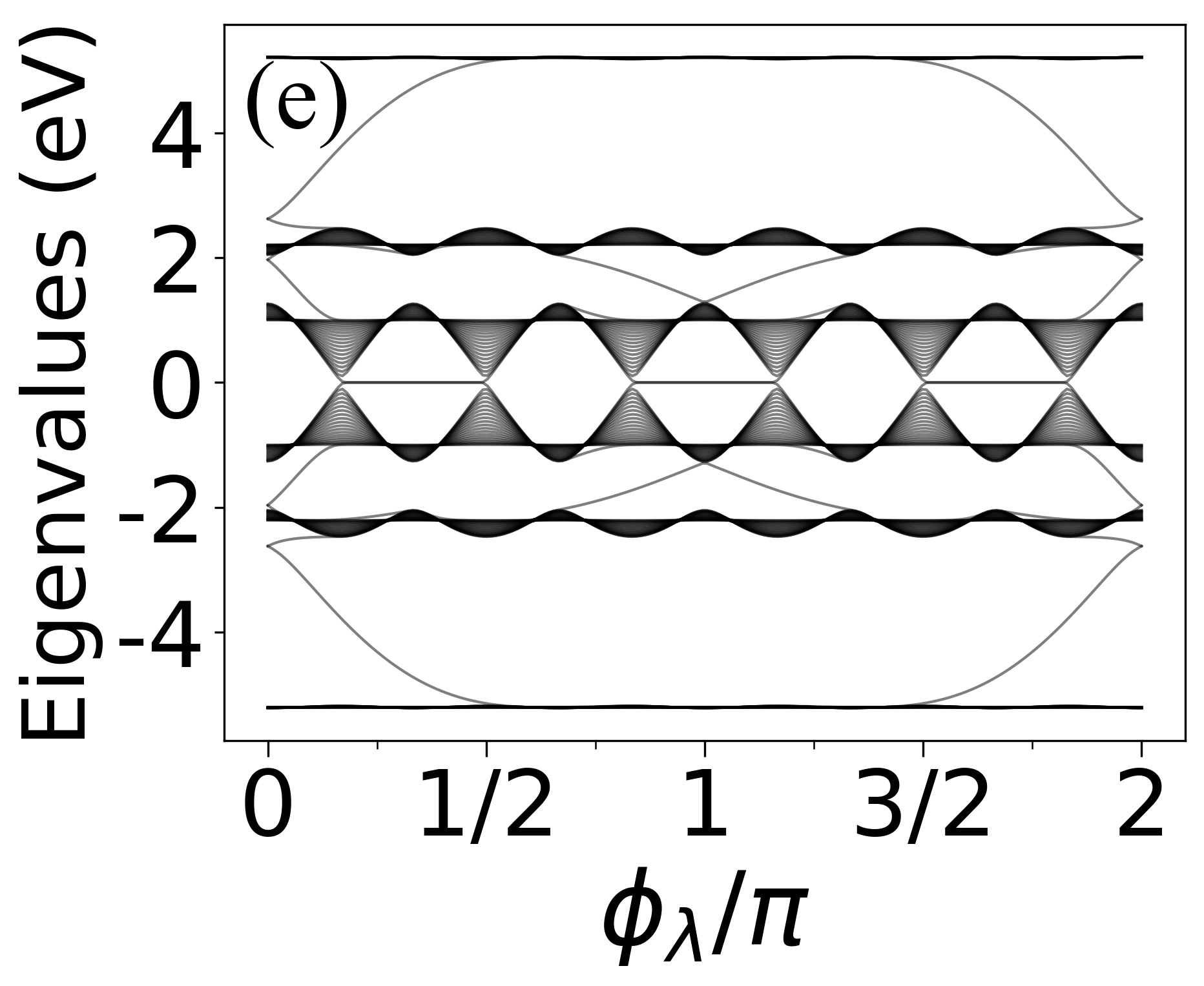}
    \end{subfigure}
    \caption{Energy spectra of the same AAH chain in Fig.~\ref{fig1}, plotted as a function of $\phi_{\lambda}/\pi$ for five representative values of modulation strength: (a) $\delta_t=1.0$, (b) $\delta_t=W_{c1}=2/\sqrt(3)$, (c) $\delta_t=1.5$, (d) $\delta_t=W_{c2}=2$, and  $\delta_t=3.0$.}
    \label{fig2}
\end{figure}

Before addressing the transport characteristics, it is instructive to examine the eigenvalue spectrum of the isolated AAH chain. Figure~\ref{fig1} illustrates the eigenspectra for the commensurate case $b=1/6$ at two representative modulation phases: $\phi_{\lambda}=0$ and $\phi_{\lambda}=\pi/2$. Several salient features emerge. For both phase configurations, the eigenvalues remain nearly identical up to $\delta_t \approx 1$, apart from the deviations of the presence of localized edge states at zero and non-zero energies. Beyond this point, the bulk energy bands, except for the outermost ones, exhibit opposite dispersive behaviors between the two phase configurations in Fig~\ref{fig1}(a) and (b). This contrast is not confined to these specific phase values but repeats periodically with $\phi_{\lambda}$, as demonstrated in Fig.~\ref{fig2}. A consequence of this inversion is that for $\phi_{\lambda}=\pi/2$, only five distinct energy bands appear, whereas for $\phi_{\lambda}=0$, six bands are resolved except at the crossing points.\vspace{0.15cm}

A particularly important observation in Fig.~\ref{fig1}(a) is the closing and reopening of the central energy band gap at the critical modulation amplitude $\delta_t=W_{c1}$ (say) accompanied by the disappearance of the zero-energy edge state; a signature of a topological phase transition. In contrast, for $\phi_{\lambda}=\pi/2$ (Fig.~\ref{fig1}(b)), the central bands undergo pronounced narrowing followed by re-broadening as $\delta_t$ increases. Despite these differences, both configurations share a common behavior: an additional gap closing  event between the central and adjacent bands around $\delta_t= W_{c2}=2$, where QH edge states merge into bulk continuum only for $\phi_{\lambda}=\pi/2$ value. Such behavior is generic  for all commensurate cases with $b=1/(2q)$ and larger $q$. These observations indicate that the AAH phase $\phi_{\lambda}$ plays a crucial role in governing the evolution of energy eigenvalues once $\delta_t > 1$. To better understand this dependence, we next examine the variation of eigenvalues with respect to $\phi_{\lambda}$ across its full range $\phi_{\lambda} \in [0, 2\pi]$.\vspace{0.15cm}

At $\phi_{\lambda}=0$, the central energy bands broaden and overlap at $\delta_t=W_{c1}$, while at $\phi_{\lambda}=\pi/2$, they collapse into a single energy level at $E=0$ eV. This complementary behavior enables an analytical estimation of $W_{c1}$. For $\phi_{\lambda}=\pi/2$, the central flat dispersion implies that at least one hopping amplitude vanishes, since the AAH Hamiltonian includes only off-diagonal terms.
The modulated hopping amplitudes are given by  
\begin{equation}
t_i= t_0(1+ W_{c1} \cos(\frac{\pi}{3}  i + \pi/2))
\label{equn}
\end{equation}
From this relation, one find that $t_1=t_2=t_4=t_5=0$ when $|W_{c1}|=\frac{2}{\sqrt{3}}$, while $t_3=t_6=t_0$ remain finite, independent of $W_{c1}$. Thus, the first topological transition corresponding to the complete disappearance of the zero-energy edge state occurs precisely at $W_{c1}=\frac{2}{\sqrt{3}}$. 

\begin{figure}[ht]
    \centering
    \begin{subfigure}[t]{0.23\textwidth}
        \includegraphics[width=\textwidth]{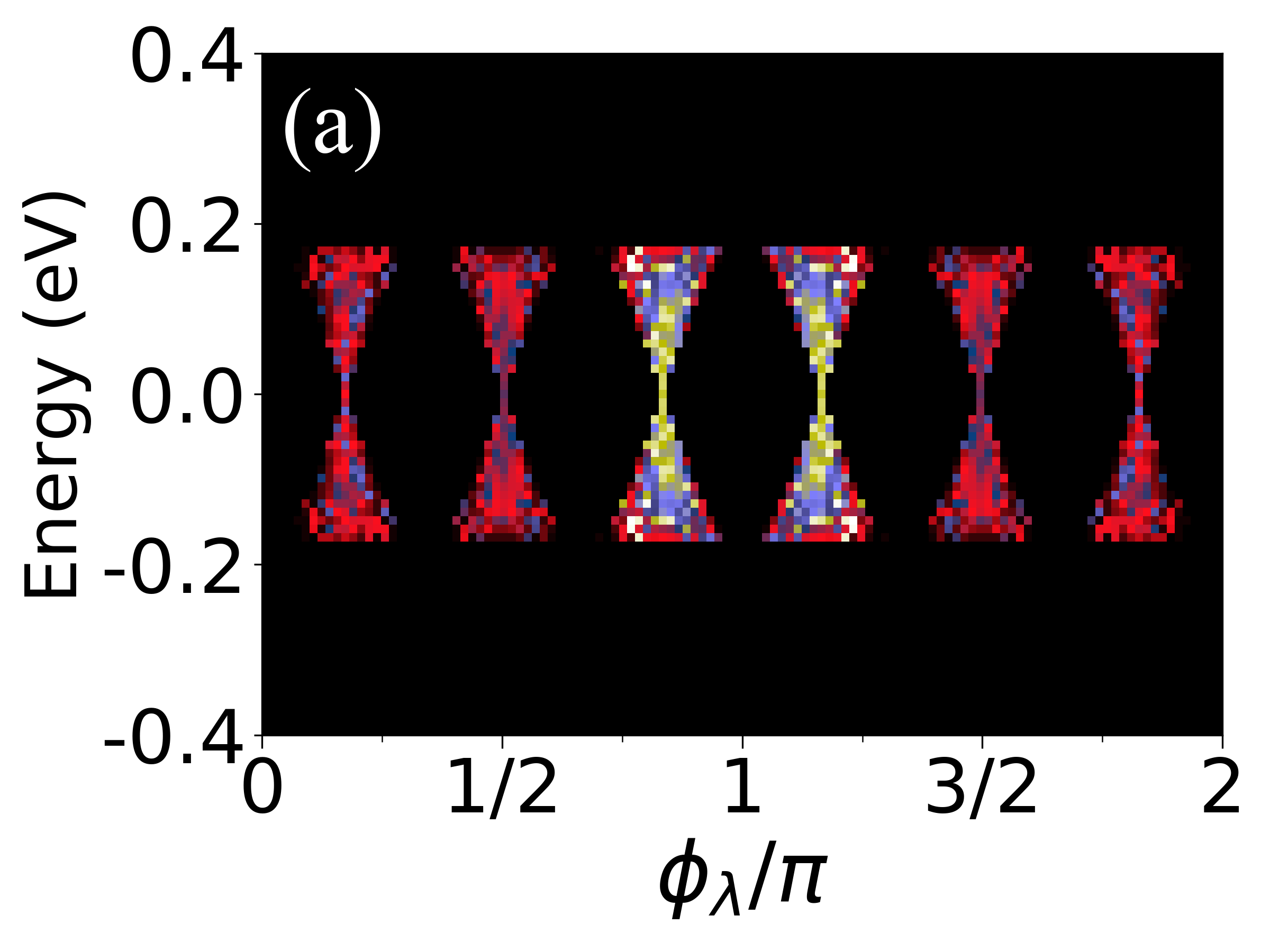}
    \end{subfigure}
    \begin{subfigure}[t]{0.23\textwidth}
        \includegraphics[width=\textwidth]{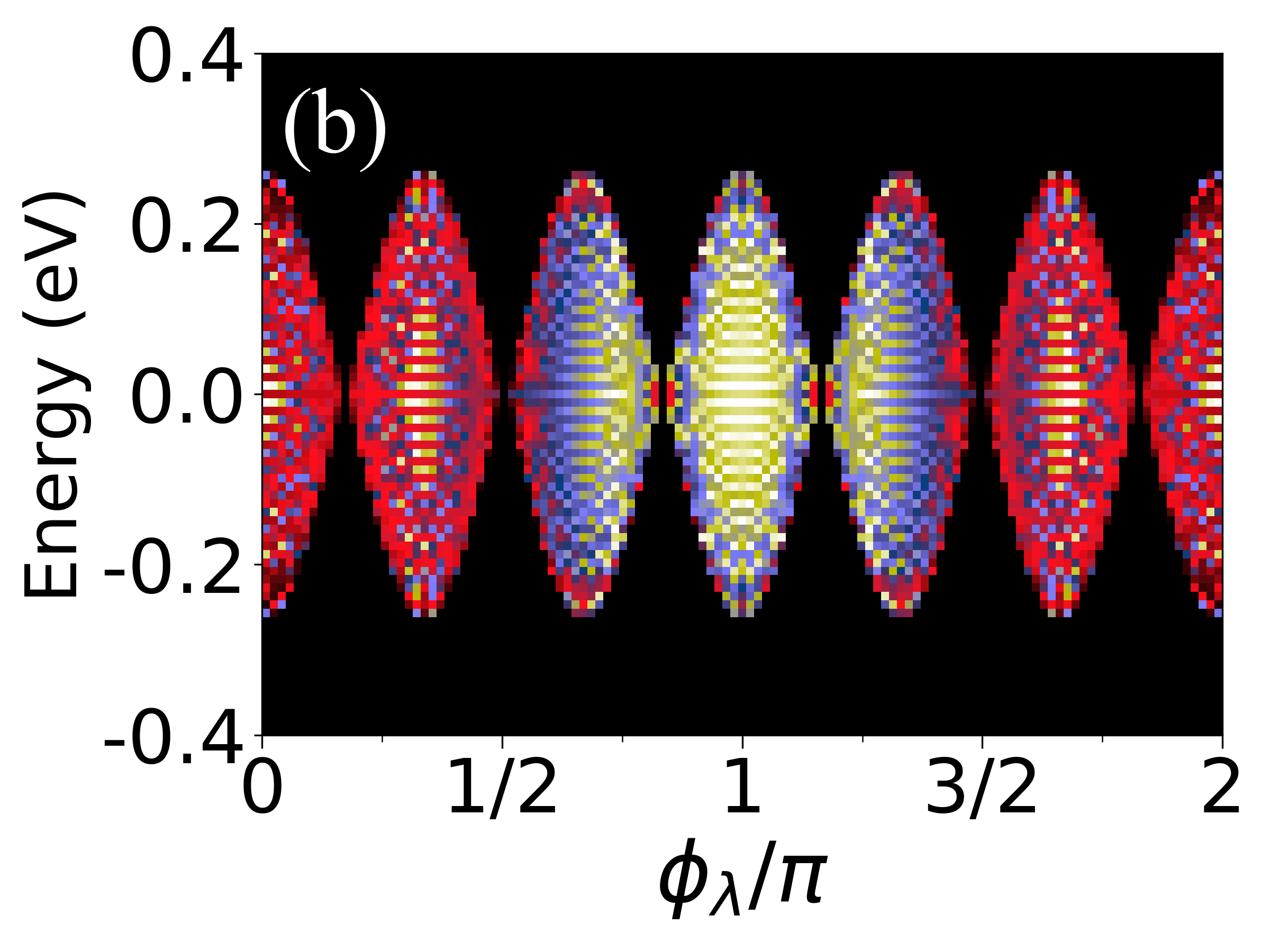}
    \end{subfigure}
    \begin{subfigure}[t]{0.23\textwidth}
        \includegraphics[width=\textwidth]{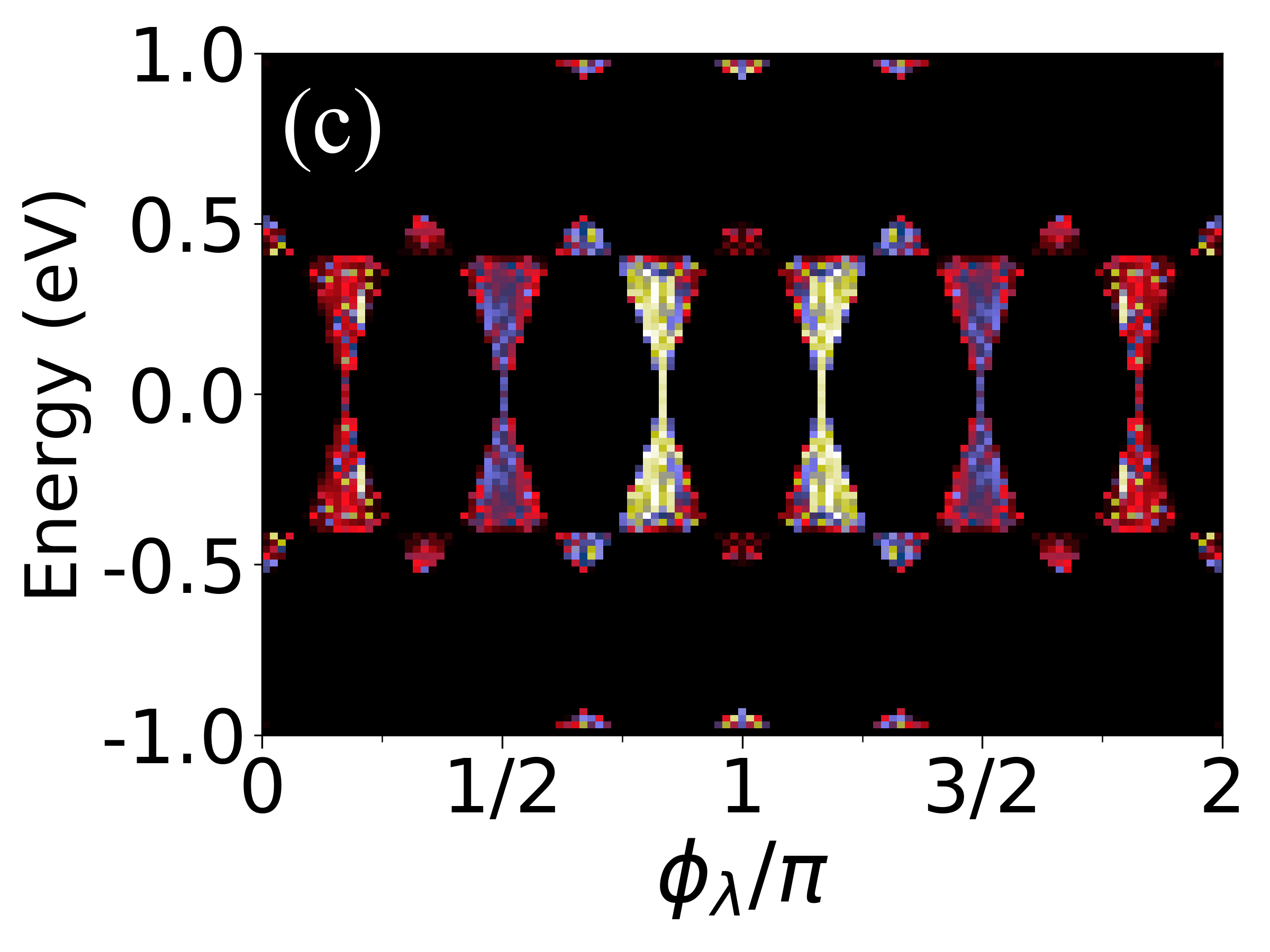}
    \end{subfigure}
    \begin{subfigure}[t]{0.23\textwidth}
        \includegraphics[width=\textwidth]{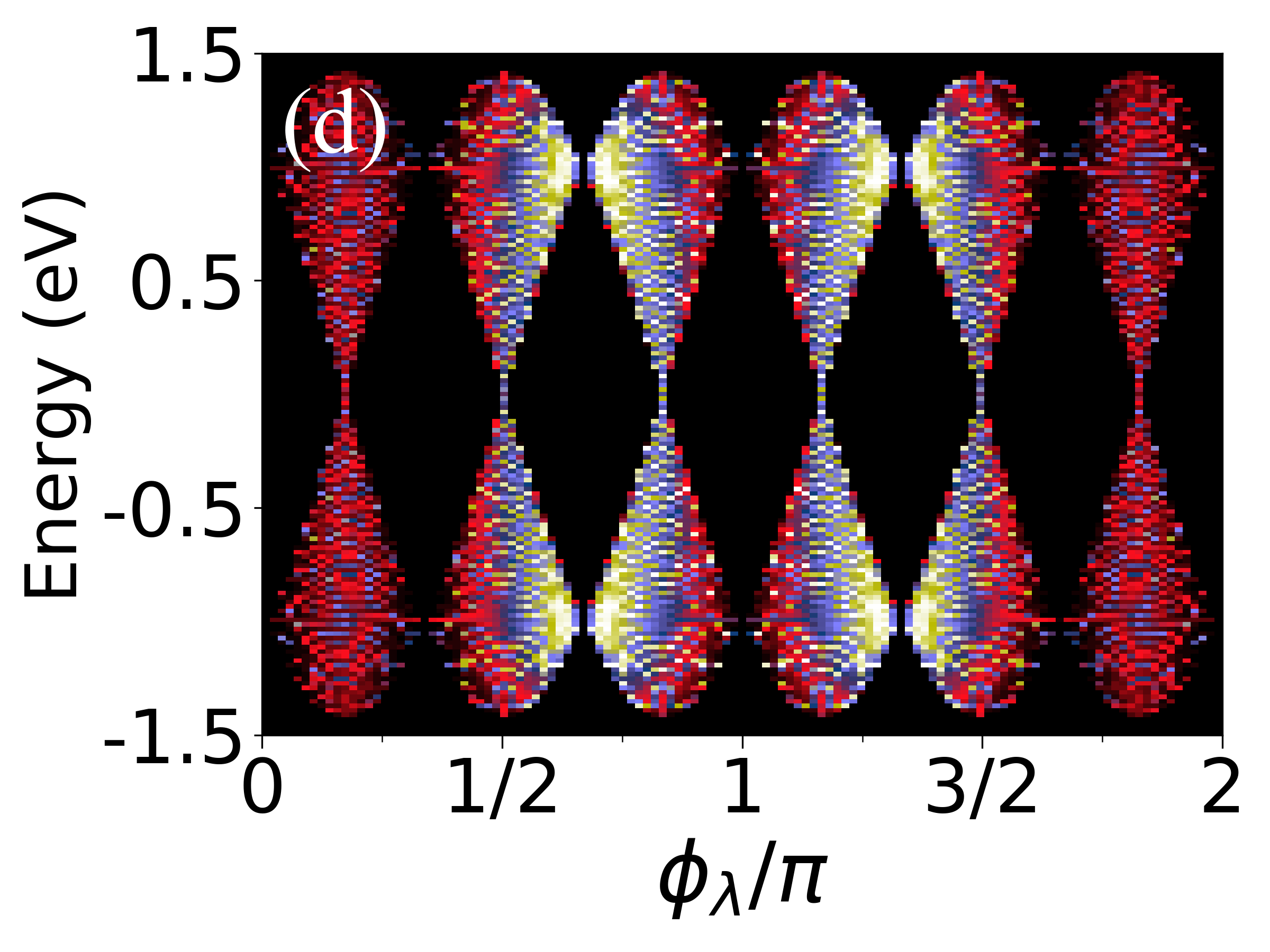}
    \end{subfigure}
    \begin{subfigure}[t]{0.23\textwidth}
        \includegraphics[width=\textwidth]{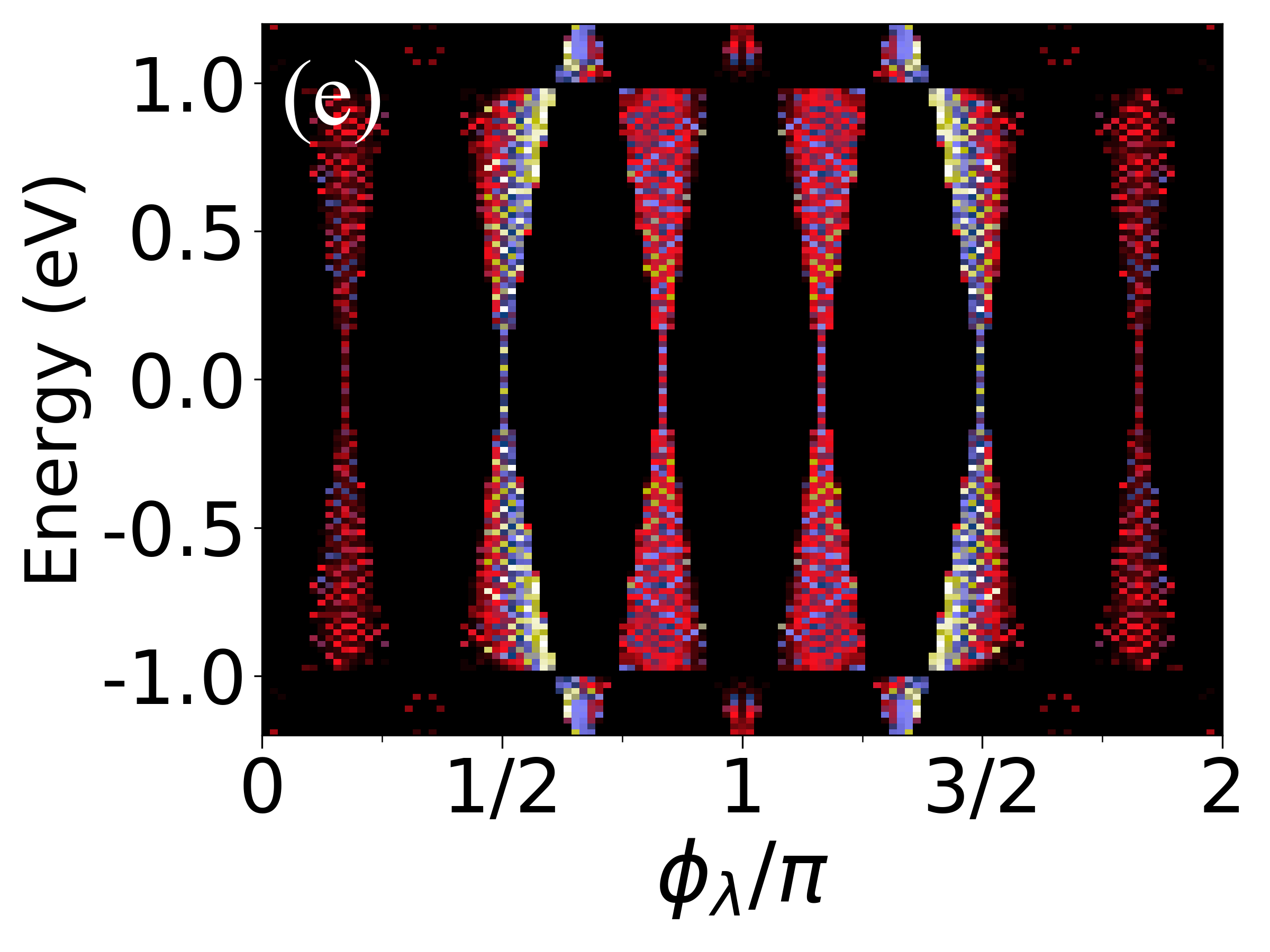}
    \end{subfigure}
    \begin{subfigure}[t]{0.05\textwidth}
        \includegraphics[width=\textwidth]{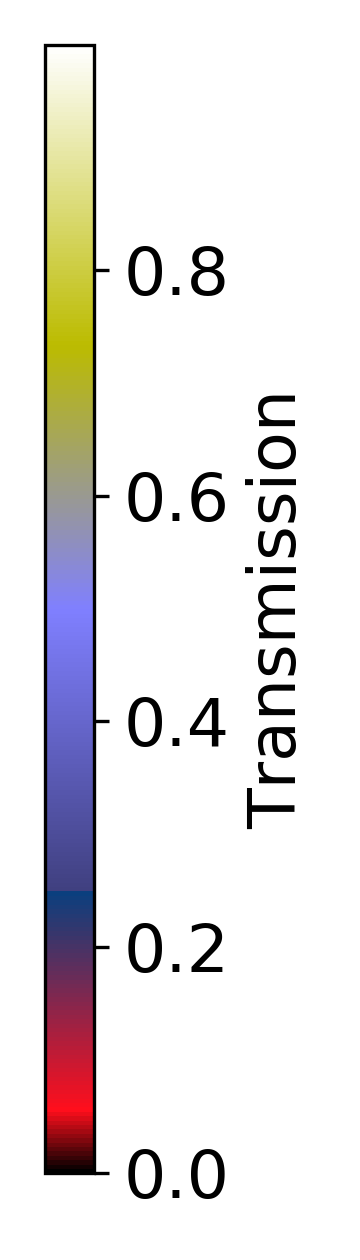}
    \end{subfigure}
    \caption{Simultaneous variation of transmission probability as a function of incoming electronic energy (in eV, vertical axis) and AAH phase $\phi_{\lambda}/\pi$ (horizontal axis) for the identical configurations as in Fig.~\ref{fig2}. The color scale represents the magnitude of transmission.}
    \label{fig3}
\end{figure}

\begin{figure}[ht]
    \centering
    \begin{subfigure}[t]{0.23\textwidth}
        \includegraphics[width=\textwidth]{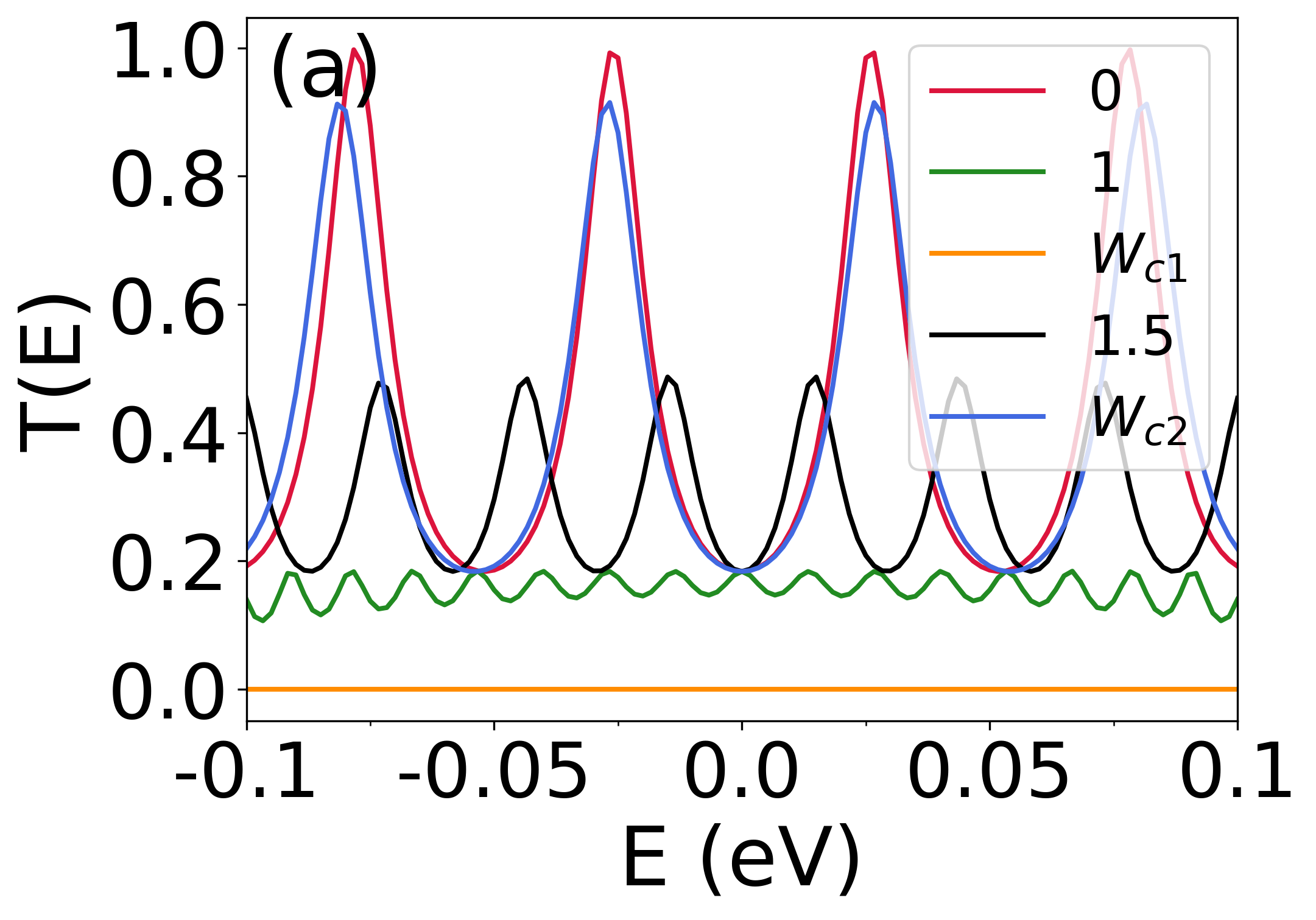}
     \end{subfigure}
    \begin{subfigure}[t]{0.23\textwidth}
        \includegraphics[width=\textwidth]{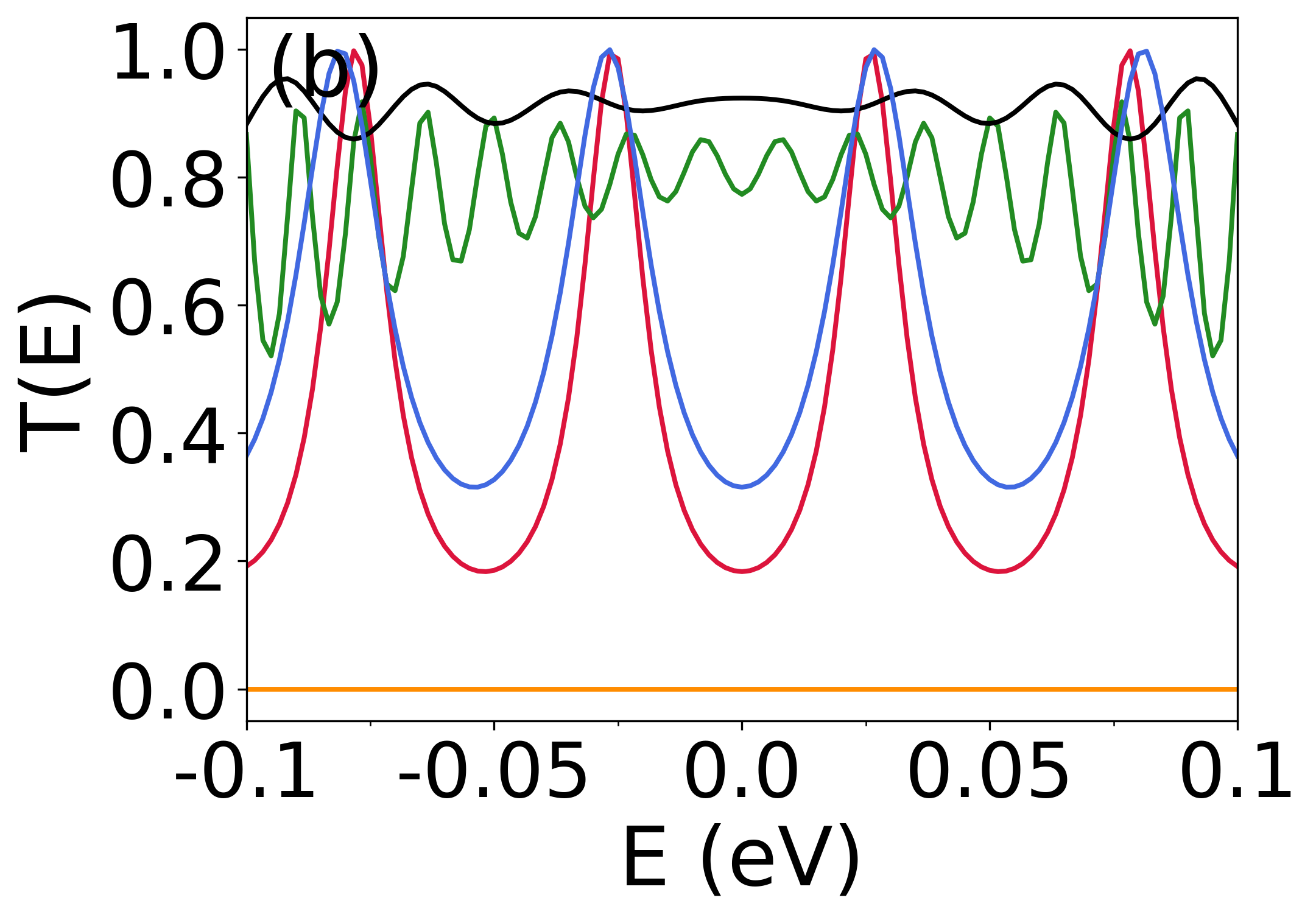}
    \end{subfigure}
    \caption{Comparison of energy-resolved transmission between the adjacent Diract points $\phi_{\lambda}=\pi/2$ in (a) and $\phi_{\lambda}=5\pi/6$ in (b) for several strengths of $\delta_t$. Results correspond to $\delta_t = 0$ (red), $1$ (green), $W_{c1}$ (orange), $1.5$ (black) and $W_{c2}$ (blue). A narrow energy window is used to capture the detailed features around the Dirac points.}
    \label{fig4}
\end{figure}
\subsection{Dependence on the modulation phase $\phi_{\lambda}$}
Figure~\ref{fig2} presents the eigenvalue spectra as functions of the AAH phase $\phi_{\lambda}$ for five representative modulation strengths $\delta_t$. A general trend observed across all cases is that the interchange of the zero-energy edge modes in Fig.~\ref{fig2}(a) and (c) with respect to $\phi_{\lambda}$ once the central bands merge at $\delta_t=W_{c1}$ in Fig.\ref{fig2}(b). This inversion persists for higher transition points such as $\delta_t= W_{c2}$ in Fig.~\ref{fig2}(d), until central bands remerge. Position of zero energy states with respect to both modulation parameters is fixed through the relation $ \left| \dfrac{\prod t_{\mathrm{odd}}}{\prod t_{\mathrm{even}}} \right|< 1$ for any generalised `q' value\cite{MB1}, here it is $ \left| \dfrac{ t_1 t_3 t_5}{ t_2 t_4 t_6} \right|< 1$.   
As stated in the previous subsection, a careful observation shows that contrasting gapped and gapless bulk bands around $E=0$ eV follows a characteristic interval of $\pi/(2q) $ in $\phi_{\lambda}$, a feature that remains valid for any commensurate modulation $b=1/(2q)$. \vspace{0.2cm}

In addition to the central zero-energy modes, QH edge states emerge within the finite energy gaps separating the bulk bands. These localized edge modes hybridize with the bulk continuum at $\delta_t=W_{c2}$ in Fig.~\ref{fig2}(d), leading to band-gap closure and subsequent reopening in Fig.~\ref{fig2}(e), consistent with the hallmark of a topological phase transition. A natural question, therefore, arises: \textit{how do these gap-closing and reopening processes influence the transport behavior of the system?} This aspect will be explored with detailed discussion in the following paragraphs.\vspace{0.2cm}

We now turn to the transport characteristics of the commensurate off-diagonal AAH chain, focusing on the central two bulk band region as a function of AAH phase. Figure~\ref{fig3} shows the simultaneous variation of total transmission as a function of incoming electronic energy along $y$-axis and AAH phase $\phi_{\lambda}/\pi$ along $x$ axis for the identical cases of Fig.~\ref{fig2}. A moderate to high transmission is consistently observed around the $\phi_{\lambda}/\pi=1$ region for all specified $\delta_t$ range, seen from each subfigure of Fig~\ref{fig3}. This picture is true for all commensurate cases including $b=1/(2q)$, provided the total number of lattice sites equals an integer multiple of $2q$ (i.e., $N=2qn$). This feature has been verified numerically for multiple $q$-values (not shown here).\vspace{0.15cm}

This central high-transmission zone coincides with the parameter regime where the product of all nearest-neighbor hoppings reaches a maximum with respect to the modulation phase $\phi_{\lambda}$, signifying constructive quantum interference among the extended Bloch states. In contrast, at the transition point $\delta_t=W_{c1}$, where several hopping amplitudes vanish at Dirac points (exist for $\delta_t \neq W_{c1}$), the transmission vanishes. This high-transmission zone is also compatible with those $\phi_{\lambda}$ regions for which QH edge states penetrate into bulk bands, as evidenced by the correspondence between the respective subfigures in Figs.~\ref{fig2} and \ref{fig3}.  \vspace{0.15cm}

As the modulation amplitude $\delta_t$ increases, the high-transmission region gradually shifts away from the central $\phi_{\lambda}$ values. This shift correlates closely with the movement of the zero-energy modes as well as the presence of QH states. For instance, at $\delta_t=1$, the zero-energy state is absent within the central gap ($\phi_{\lambda} \in (\frac{5\pi}{6}, \frac{7\pi}{6})$ in Fig.\ref{fig2}(a)), and the transmission maximum appears only in the hourglass-like central bulk bands shown in Fig.\ref{fig3}(a). When $\delta_t=1.5$, the zero-energy state appears within the same bulk bands by swapping its position with respect to $\delta_t=1$ case, and the transmission speards outward w.r.t. $\phi_{\lambda}/\pi$ as represented in Fig.\ref{fig3}(c). With further increase of $\delta_t$, the transmission region shifts to the neighboring plaquette in Fig.\ref{fig3}(e), leaving the central bands almost insulating. Transmission at the transitions (i.e. $\delta_t=2/\sqrt{3}$ in Fig.~\ref{fig3}(b) and $2$ in Fig.~\ref{fig3}(d)) show some interesting features from the remaining ones, like almost maximum value of electron transmission `$1$' and their presence in different energy and $\phi_{\lambda}$ regions. Besides these, another factor is $T(E) = 0$ at $E=0$, except $\delta_t=W_{c1}$ and Dirac points, indicating the presence of localised zero energy modes. Thorough investigation shows at energy crossing regions (such as $\phi_{\lambda}=\pi/2$ in Fig.\ref{fig4}(a) and $5\pi/6$ in Fig.\ref{fig4}(b)) close to $E=0$ eV, moderate transmission exists only when they are in the high conducting zones. Value of $T(E)$ for some $\delta_t$ exceeds the unmodulated $\delta_t=0$ case. This non-trivial phenomena is clearly presented in Fig.\ref{fig4}(b). \vspace{0.2cm}

Another observation reveals that, the junction points between neighboring hourglass-like plaquettes, like $\phi_{\lambda}=0$, $\pi/3$ always exhibit vanishing transmission in each subfigure of Fig.~\ref{fig3} except in Fig.~\ref{fig3}(b) i.e., first transition point. This corresponds to the fact that any one of the corresponding six hopping amplitudes effectively vanish, suppressing coherent transport across those boundaries. Thus, by tuning the Fermi energy and the AAH phase, one can drive the system from a high to a low-transmission state (or vice versa) for a broad range of $\delta_t$, offering an effective knob to control conductance in topological AAH lattices. \vspace{0.2cm}
\begin{figure}[ht]
    \centering
    \begin{subfigure}[t]{0.23\textwidth}
        \includegraphics[width=\textwidth]{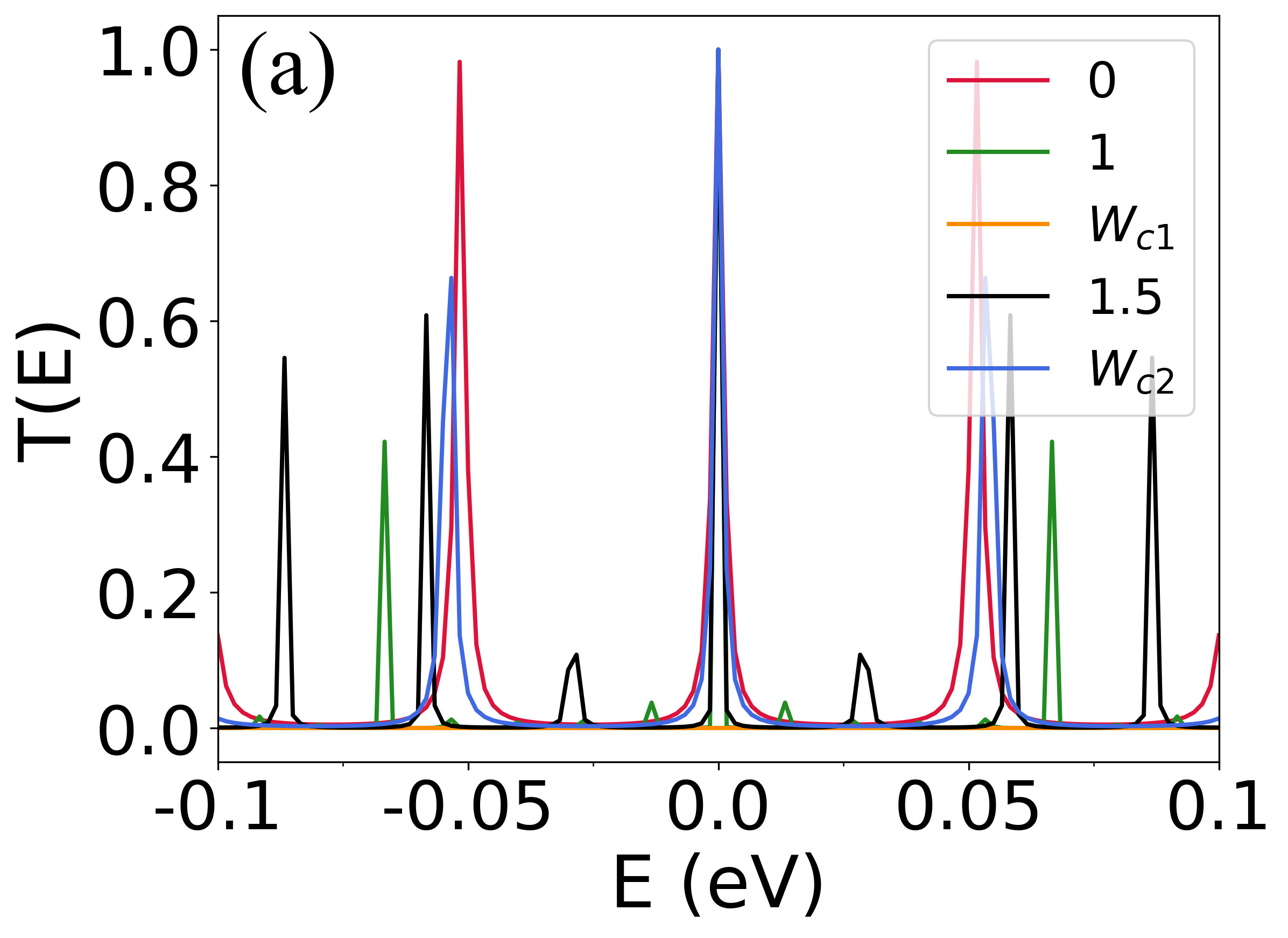}
     \end{subfigure}
    \begin{subfigure}[t]{0.23\textwidth}
        \includegraphics[width=\textwidth]{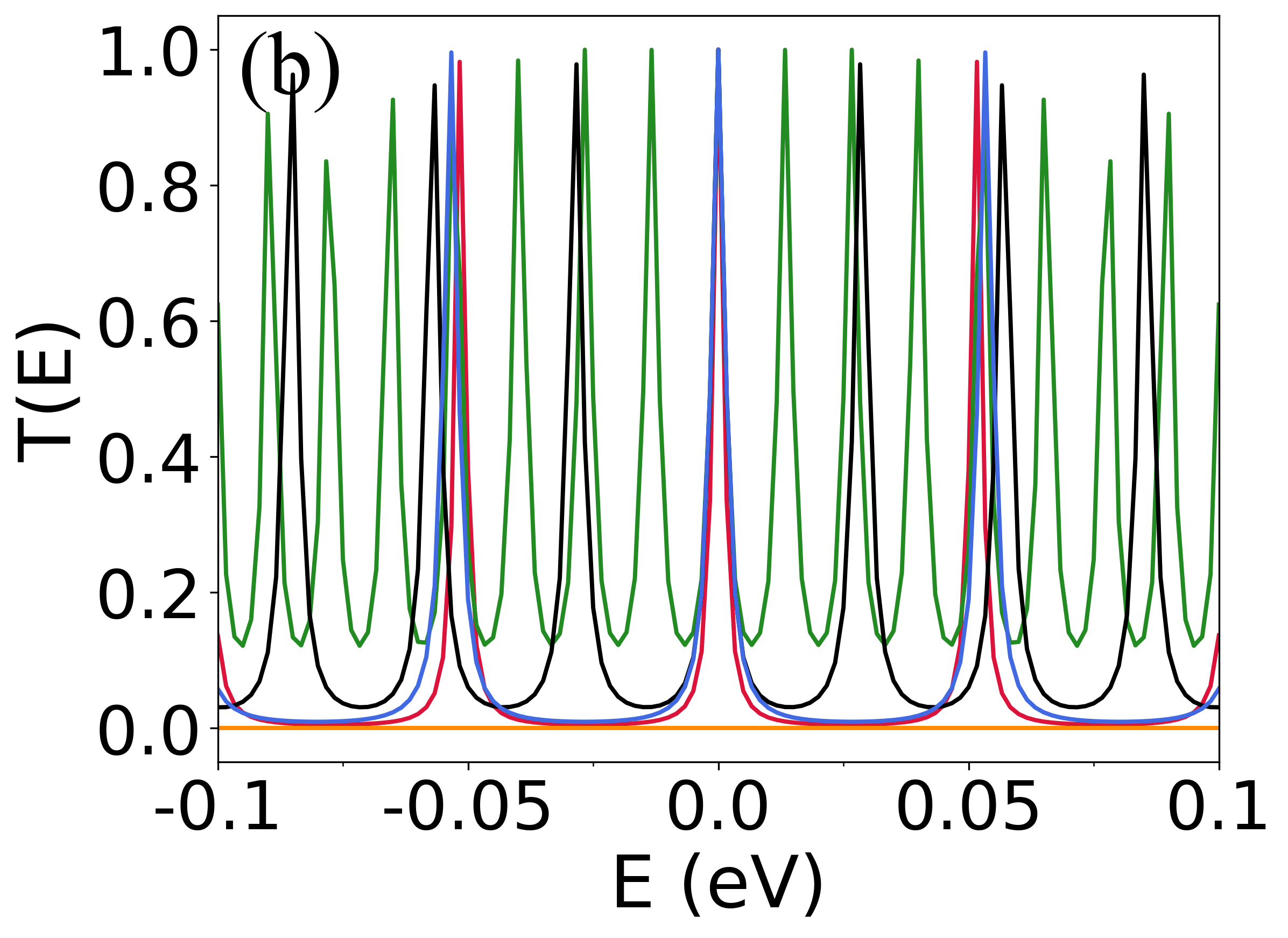}
    \end{subfigure}
     \caption{Ballistic transmission at $E=0$ eV for two adjacent Diract points $\phi_{\lambda}=\pi/6$ in (a) and $\phi_{\lambda}=\pi/2$ in (b) for several values of $\delta_t$ using identical color scheme as in Fig.~\ref{fig4}. To highlight the  robustness of zero-energy transmission $\tau_S=\tau_D$ are fixed at $0.3$ eV.}
    \label{fig5}
\end{figure}  
\subsection{Even–odd effect in lattice size}           
In addition to the intrinsic parameters of the AAH potential ($\delta_t$ and $\phi_{\lambda}$), the total number of lattice sites plays a crucial role in shaping the transport response. For commensurate modulation $b=1/(2q)$, the total length of the chain can independently vary between $2qn$ to $2q(n+1)-1$, which effectively alters the final hopping amplitude connecting the conductor to the drain lead and this phenomena left distinct characteristics in electron transmission. For our case, $b=1/6$, if number of lattice sites is varied from $6n$ to $6n\pm 5$, following are the observations. QH edge states, within the gapped bulk band regions, exhibit pronounced shifts in their positions, governed primarily by the terminating hopping amplitude. This modulation, in turn, strongly influences the emergence of different high-conductance regions with respect to $\phi_{\lambda}$. Upto Fig.~\ref{fig4}, the electron transmission is analyzed in detail for a representative chain of $N=6n=120$ sites. \vspace{0.2cm}
\begin{figure}[ht]
    \centering
    \begin{subfigure}[t]{0.23\textwidth}
        \includegraphics[width=\textwidth]{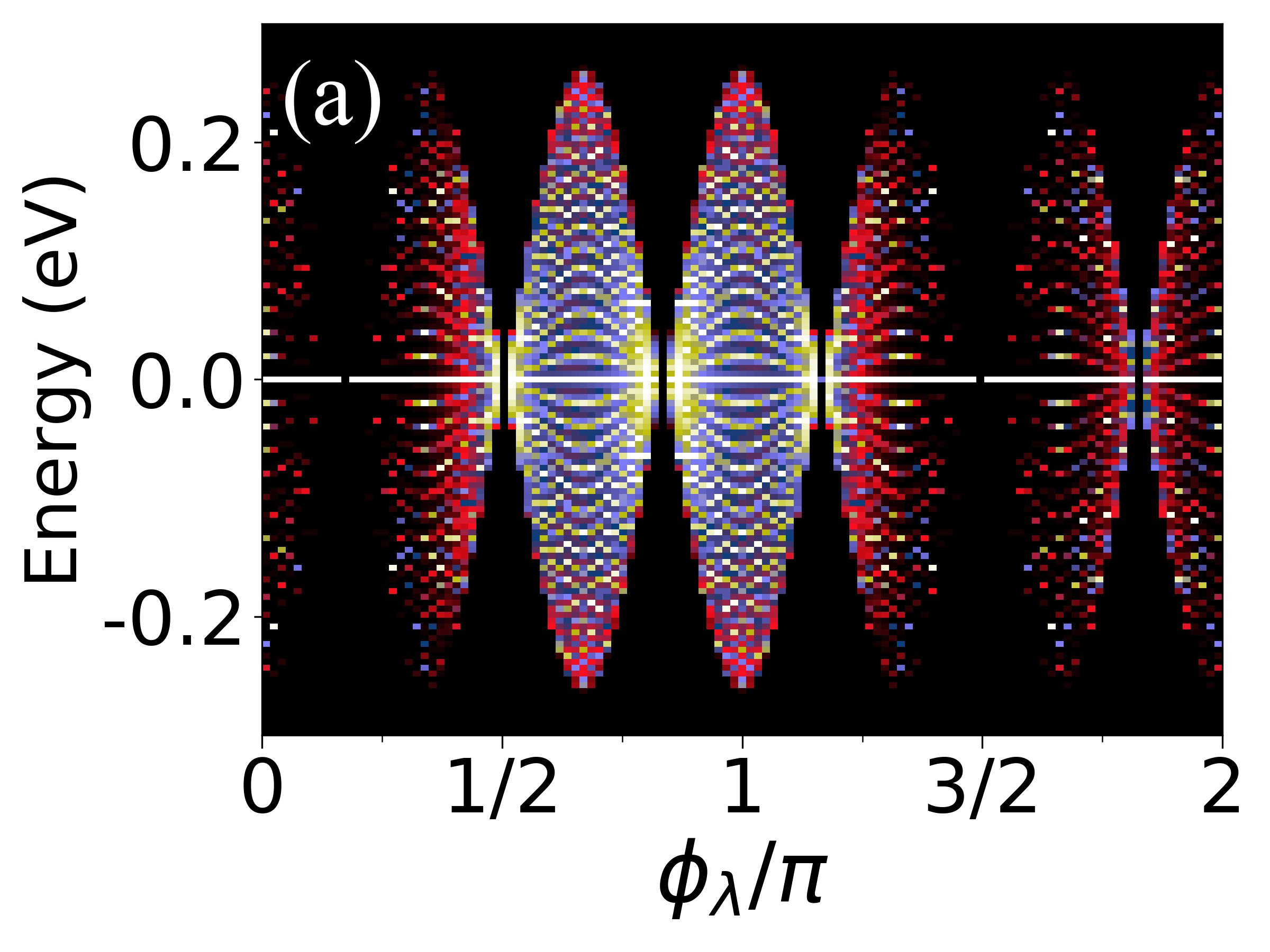}
     \end{subfigure}
    \begin{subfigure}[t]{0.23\textwidth}
        \includegraphics[width=\textwidth]{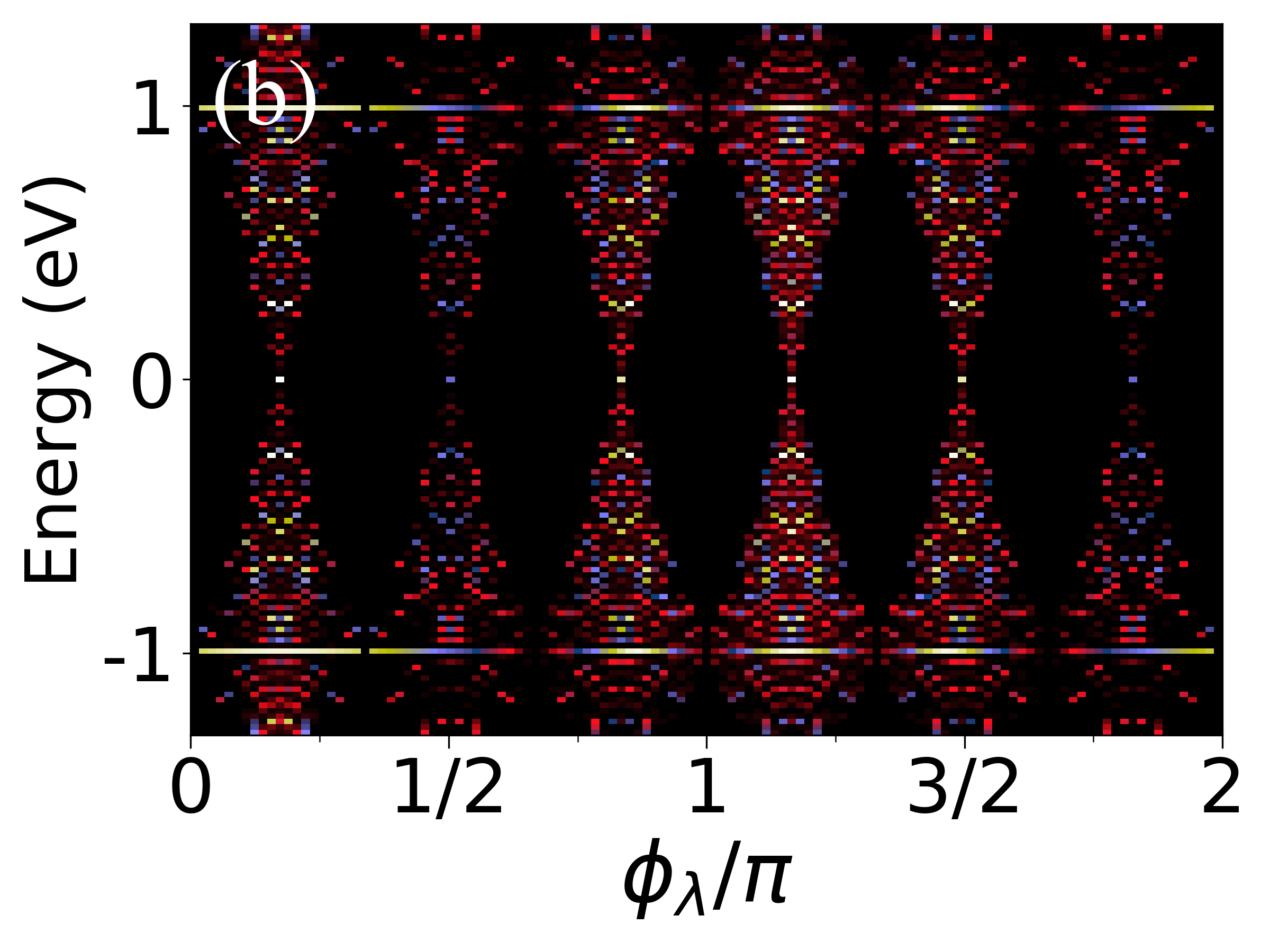}
    \end{subfigure}
    \caption{Density plots of transmission as functions of energy and the AAH phase for (a) $N=121$, $\delta_t=2/\sqrt(3)$, and (b) $N=125$, $\delta_t=2.0$ with identical couplings $\tau_S=\tau_D= 0.3$ as Fig.~\ref{fig5}. Color scale of the magnitude of transmission is identical to Fig.~\ref{fig2}.}
    \label{fig6}
\end{figure}

\begin{figure}[ht]
    \centering
    \begin{subfigure}[t]{0.23\textwidth}
        \includegraphics[width=\textwidth]{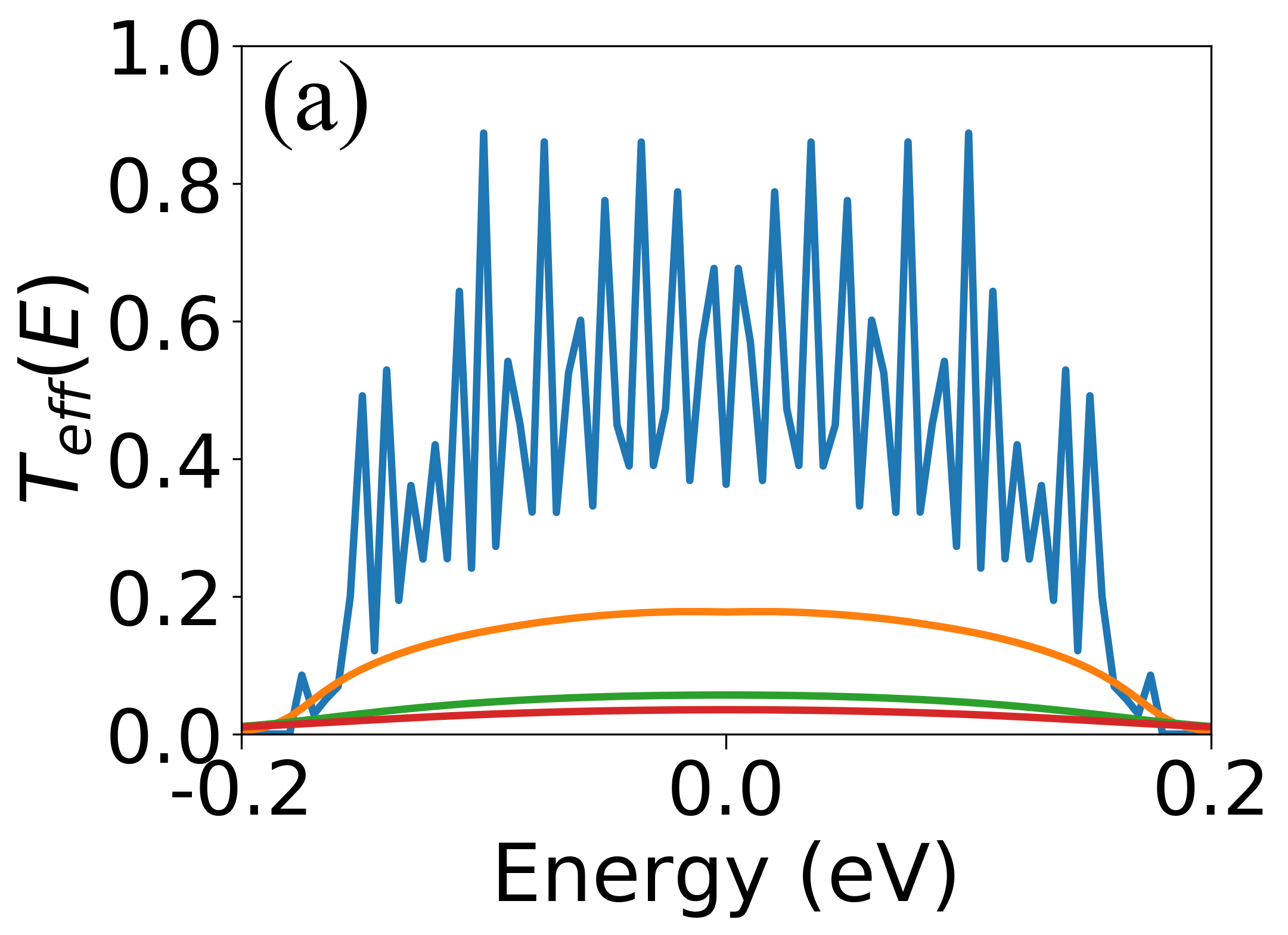}
    \end{subfigure}
    \begin{subfigure}[t]{0.23\textwidth}
        \includegraphics[width=\textwidth]{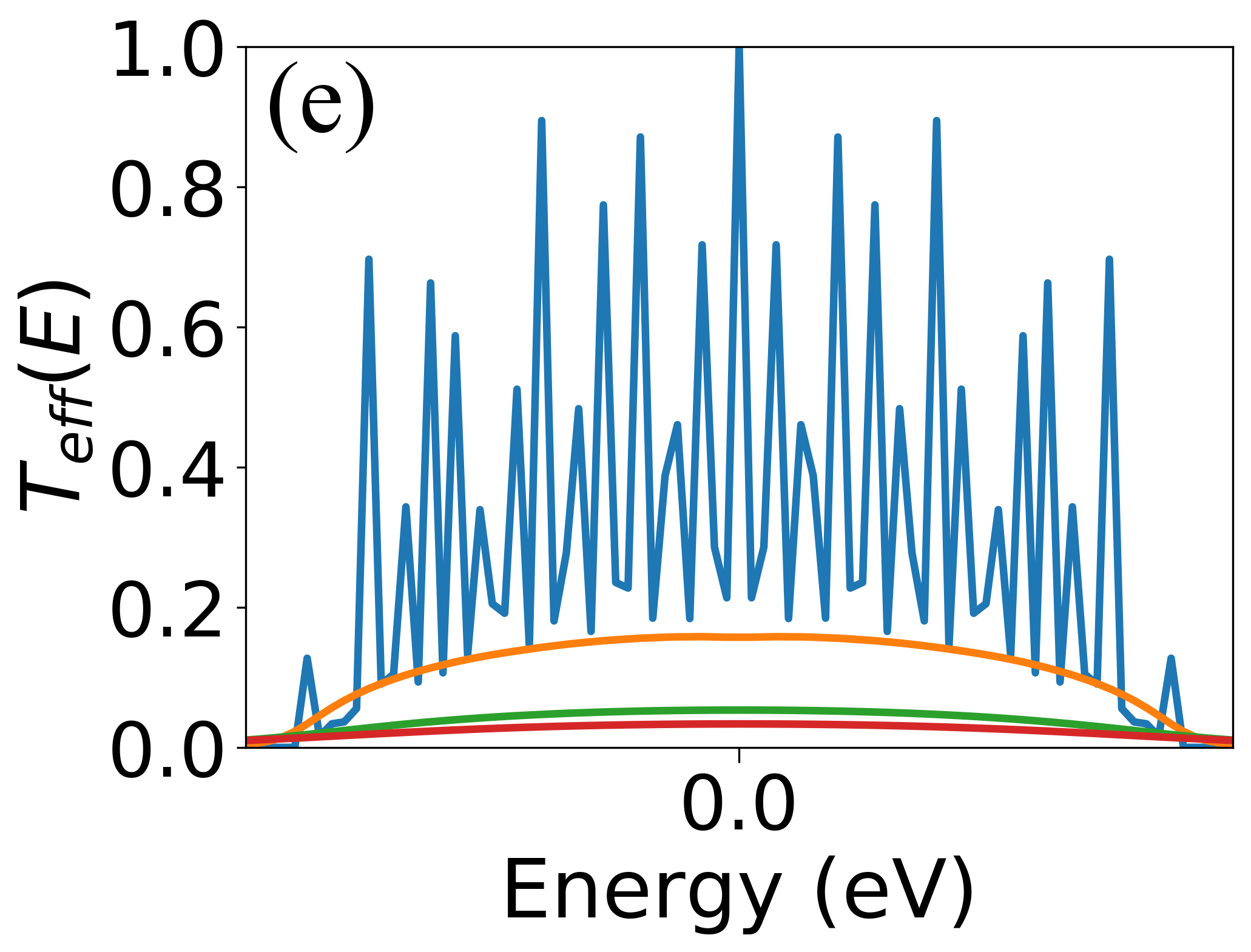}
    \end{subfigure}
    \begin{subfigure}[t]{0.23\textwidth}
        \includegraphics[width=\textwidth]{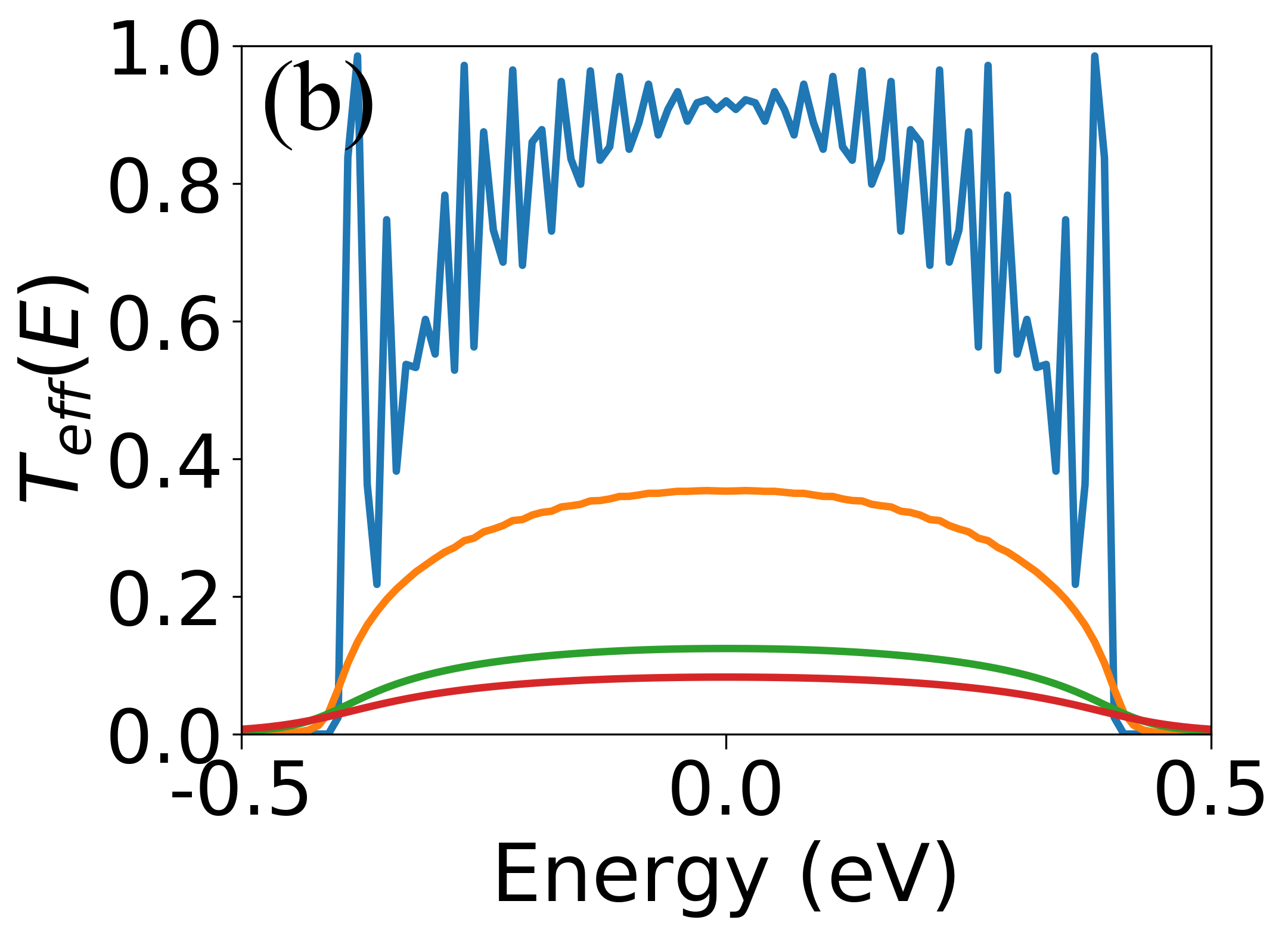}
    \end{subfigure}
    \begin{subfigure}[t]{0.23\textwidth}
        \includegraphics[width=\textwidth]{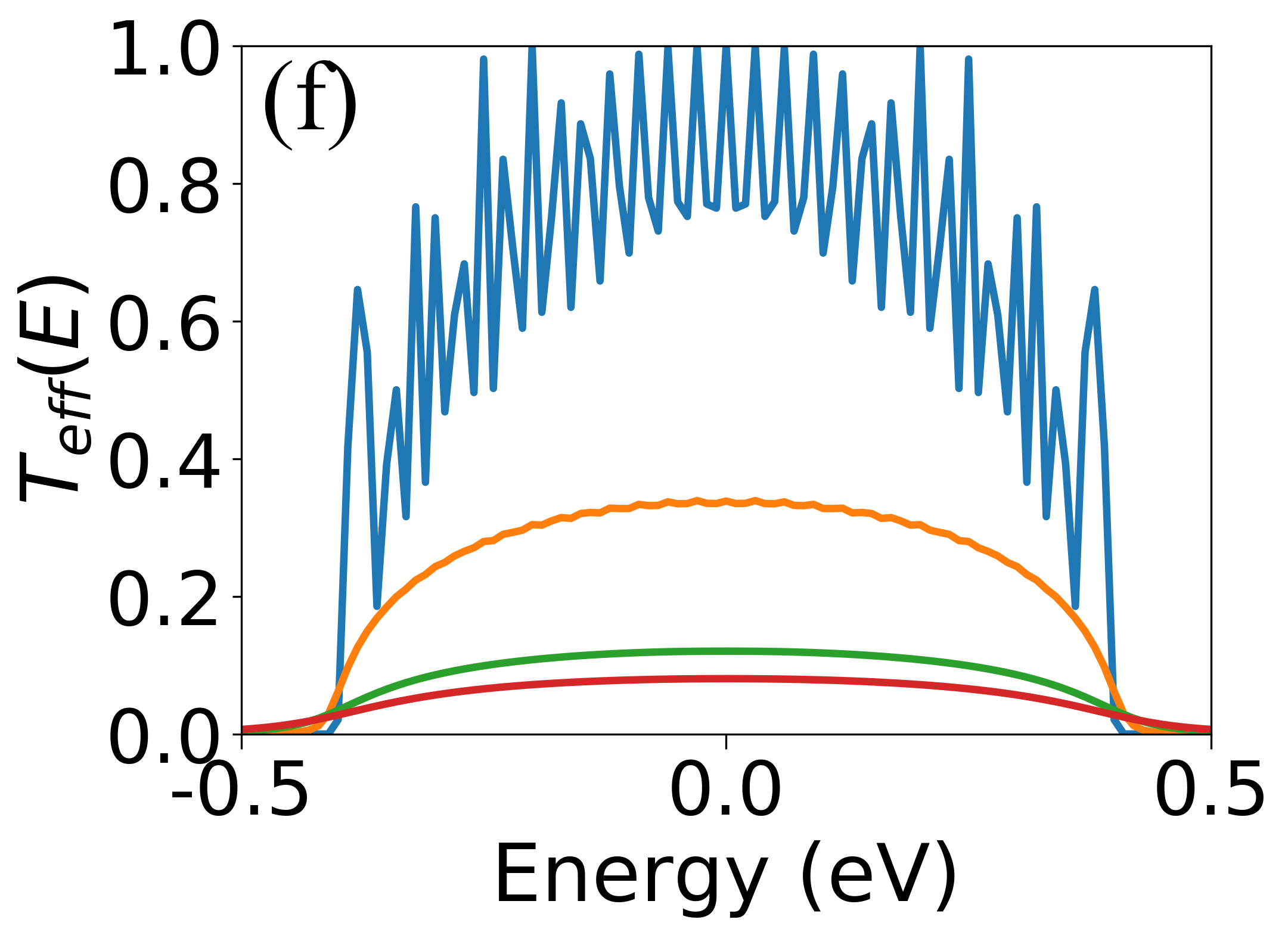}
    \end{subfigure}
    \begin{subfigure}[t]{0.23\textwidth}
        \includegraphics[width=\textwidth]{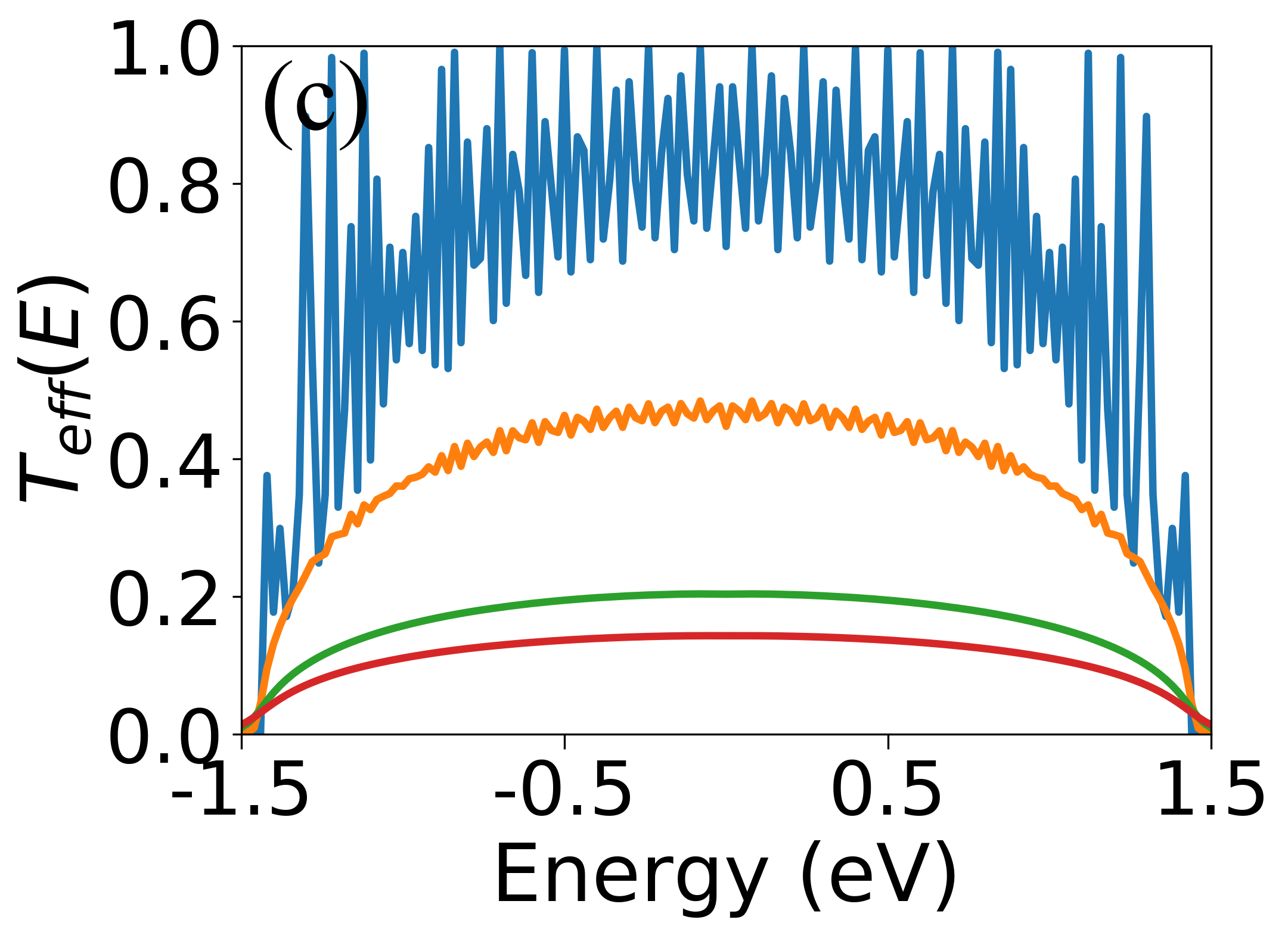}
    \end{subfigure}
       \begin{subfigure}[t]{0.23\textwidth}
        \includegraphics[width=\textwidth]{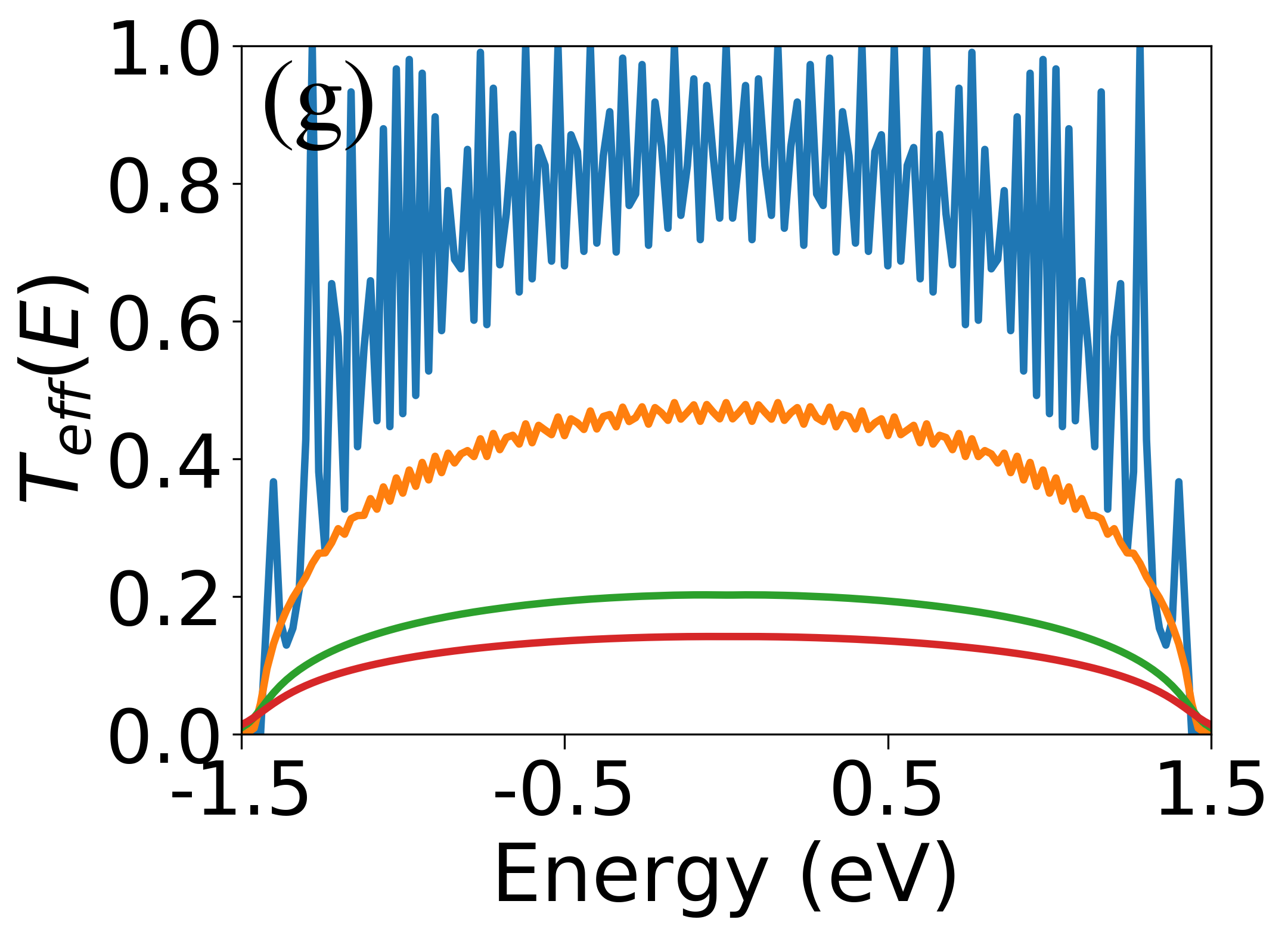}
        \end{subfigure}
            \begin{subfigure}[t]{0.23\textwidth}
        \includegraphics[width=\textwidth]{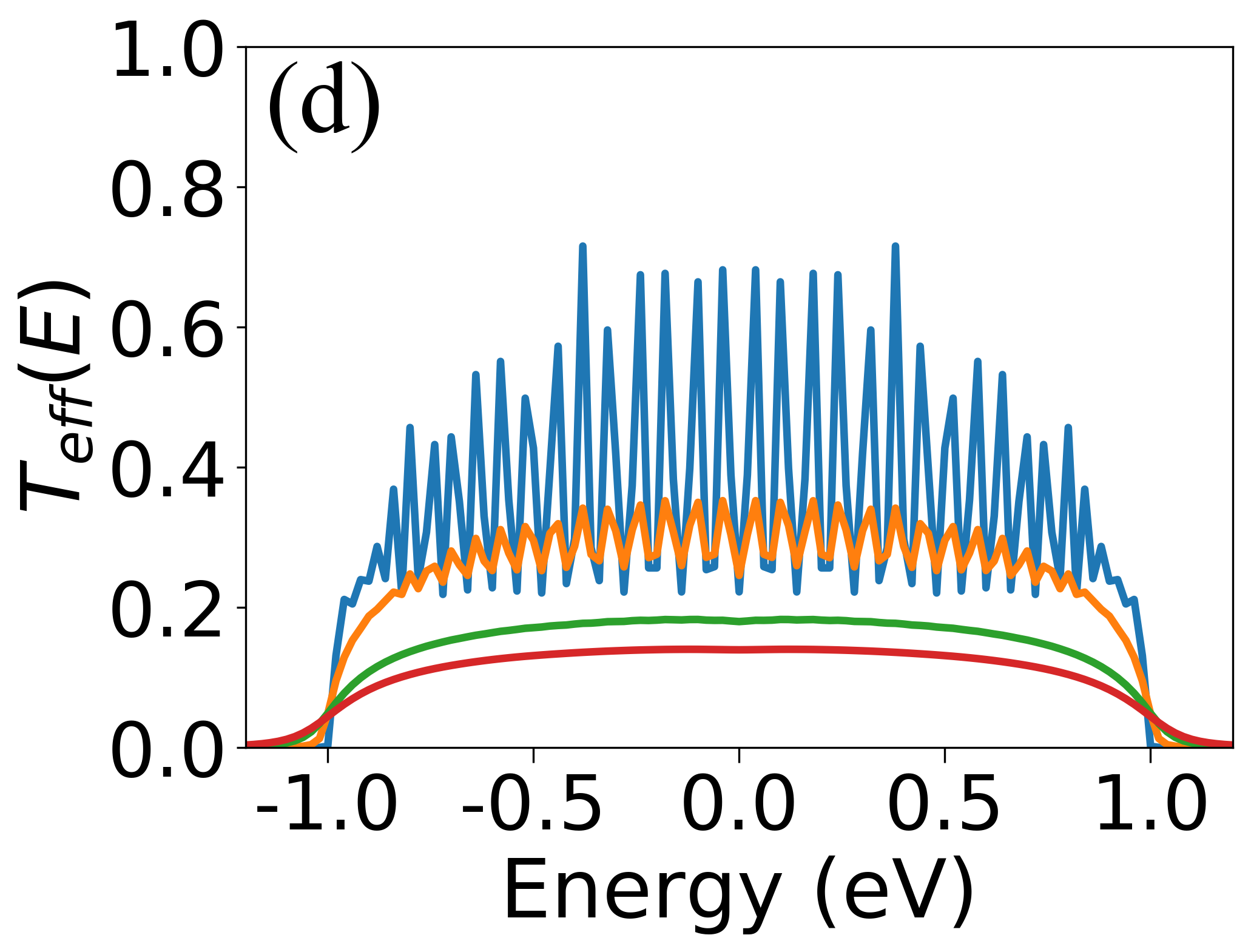}
    \end{subfigure}
       \begin{subfigure}[t]{0.23\textwidth}
        \includegraphics[width=\textwidth]{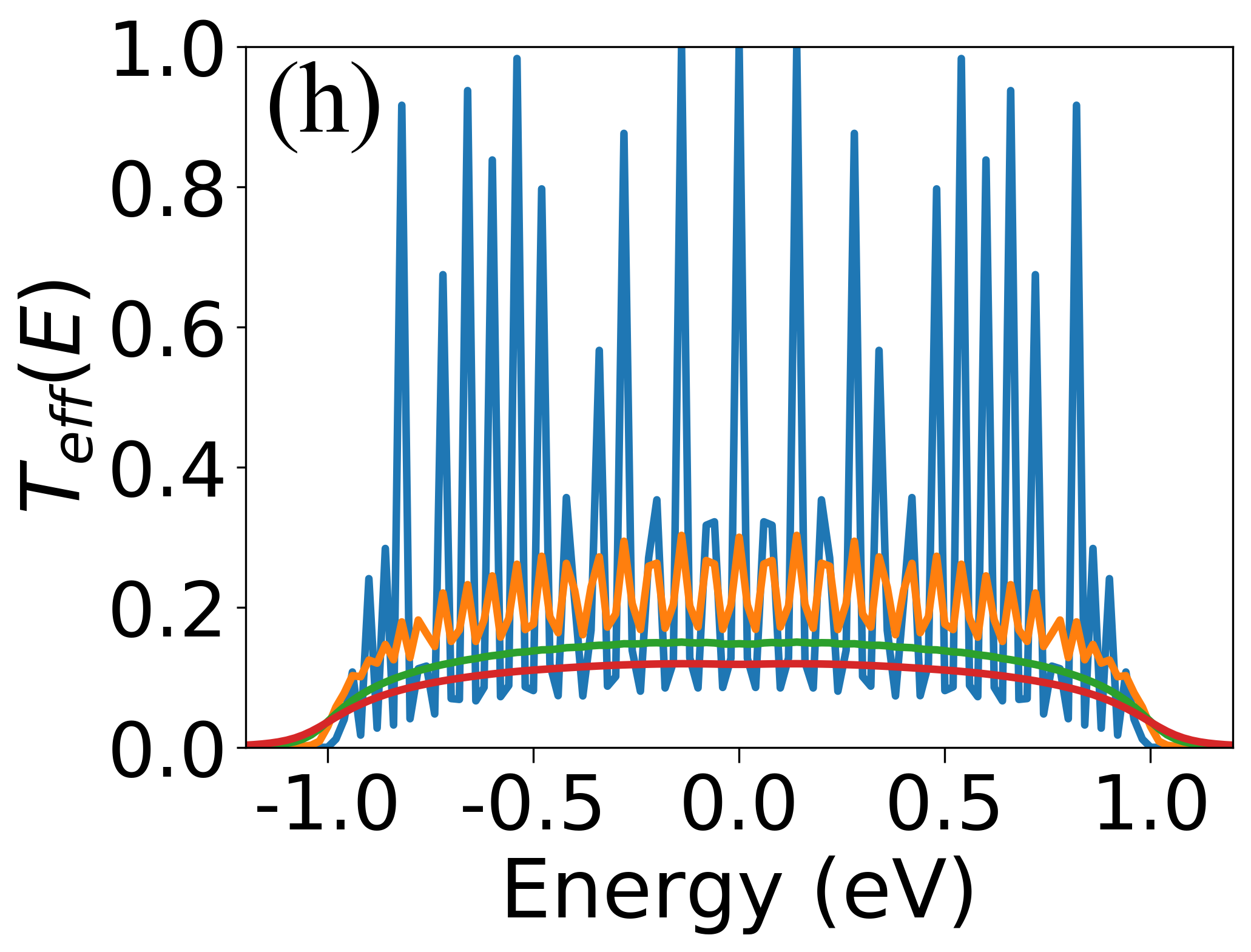}
    \end{subfigure}
    \caption{Effective transmission $T_{eff}$ as a function of electron energy for $\delta_t=1.0$ in first row, $1.5$ along second row, $2.0$ in third row and $3.0$ in last one. Results are shown for dephasing strengths $\tau_p=0$ (blue), $0.2$ (orange), $0.4$ (green), and $0.5$ (red), at $\phi_{\lambda}=5 \pi/6$, one of the Dirac points. Entire first column represents lattice size $N=120$ (a)-(d) and second one for $N=121$ (e)-(h). Lead-chain couplings are fixed at $\tau_S=\tau_D=1$ eV. Increasing $\tau_p$ leads to suppression and broadening of transmission due to inelastic scattering.}
    \label{fig7}
\end{figure}
 
Consistent with the observations reported by the paper ~\cite{SDS1}, even-odd parity dependence leads to qualitatively distinct spectral and transport characteristics. In particular, the zero-energy edge states remain pinned and robust with a certain $\phi_{\lambda}$ regime for all even-site systems, whereas in odd-site configurations, these modes persist throughout the full range of $\phi_{\lambda}$ except at Dirac points. This single zero-energy mode in odd cases arises from the breaking of chiral symmetry and is localized at either of the edges. All these descriptions are valid for any off-diagonal commensurate $b=1/(2q)$ case. \vspace{0.15cm}

We now focus on the even-odd lattice size effect, examined at three characteristic regimes: the Dirac points and the two transition points. After performing a systematic numerical analysis for lattice sizes ranging from $N=121$ to $N=125$, it comes out that $N=121$ lattice site produce robust, ballistic transport at all Dirac points and for each modulation strength $\delta_t$ considered in Figs.~\ref{fig2} and \ref{fig3}, except $\delta_t=W_{c1}$. This fact is displayed by picking two adjacent Dirac points $\phi_{\lambda}=\pi/6$ and $\pi/2$ in Fig~\ref{fig5}(a) and (b) respectively. This effect persists for any general $b=1/(2q)$ off-diagonal commensurate AAH chain with $2qn+1$ lattice sites. \vspace{0.15cm}

Remarkably, even for $\delta_t = W_{c1}$ ballistic transmission persists at $E=0$ eV throughout $\phi_{\lambda}$ range except nullified hopping cases. This robust phenomena at first transition point with $N=121$ configuration is shown in Fig.~\ref{fig6}(a). A closer inspection further reveals that at the second transition point $\delta_t= W_{c2}$, a strong transmission peak approaching ballistic magnitude emerges near $E=\pm t_0$ eV exclusively for $N=125$ (i.e., the 6n+5 configuration), as illustrated in Fig.~\ref{fig6}(b). All these phenomena associated to odd chains largely insensitive to the specific lead-molecule coupling strength and this is absent in even site chains. To show the robustness of the \textit{perfect ballistic transmission}, we have decreased the lead to conductor coupling to a minimum value $\tau_S=\tau_D=0.3$ eV for all figures in Figs~\ref{fig5} and \ref{fig6}. As mentioned in text, shifting of high-conducting zone with changing lattice size is clear from Fig.~\ref{fig6}(a).  \vspace{0.15cm}

\subsection{Inclusion of dephasing by B\"{u}ttiker probe}
The inclusion of Büttiker probes introduces inelastic scattering, which decreases the transmission magnitude but broadens the transmission spectrum due to phase randomization by the voltage probes. This behavior is well established in the  literature.~\cite{MBP2,MBP3,BP1,BP2,BP0} In particular, M. Saha $\textit{et al}$. ~\cite{BP2} demonstrated that in an incommensurate AAH lattice, coupling to Büttiker probes gives rise to a finite conductance even in the fully localized regime, with the conductance following a power-law scaling with the probe-conductor coupling strength. This scaling changes drastically from $\tau_p < 1$ to $\tau_p > 1$.\vspace{0.15cm}

To examine the impact of environmental decoherence, we evaluate the effective transmission $T_{eff}$ (Eq.\ref{equ9}) as a function of incoming electron's energy for discrete values of dephasing strength $\tau_p$,  which is varied up to $0.5$ eV, corresponding to the weak-coupling regime. Figure \ref{fig7} presents $T_{eff}$(E) for several configurations in the central gapless regime at the Dirac point $\phi_{\lambda}=5\pi/6$. The left and right columns correspond to chains with $N=120$ (even)  and $N=121$ (odd) sites, respectively, allowing a direct comparison of even-odd effects in the presence of dephasing. Each row represents a fixed modulation strength $\delta_t=1.0$, $1.5$, $2.0$ and $3.0$, respectively, from up to down.\vspace{0.15cm}

The overall difference between the even and odd lattices diminishes with increasing $\delta_t$, nearly vanishing around the second transition point $W_{c2}$ in Fig.~\ref{fig7}(c) and (g), but re-emerges beyond it. This reflects the shifting of high conducting zone with changing the final hopping in the AAH chain, as previously mentioned in Sec.\ref{sec:numresult}(C). Nature of $T_{eff}(E)$ with $\tau_p=0$ in Fig.\ref{fig7}(a)-(d) is already shown in Fig.\ref{fig3}. For finite B\"{u}ttiker probe coupling strength, inelastic scattering decreases the coherent nature of electron movement and smoothens the kink-like nature of transmission. Inclusion of dephasing enhances $T_{eff} (E)$ near the band center with increasing $\delta_t$ and persists upto $\delta_t=2$. In that case  $\tau_p=0.2$ eV shows effective transmission $T_{eff}(E)=0.4$ in the energy band center for both even and odd cases. It reflects the dephasing included system’s evolution toward a high-conducting regime with increasing modulation strength. At $\delta_t=3$ the coherent transmission is strongly suppressed across the row, yet the incoherent contribution remains appreciable and, in some energy windows in Fig.~\ref{fig7}(h) the transmission for $\tau_p=0.5$ exceeds the no-dephasing value.\vspace{0.15cm} 

Further analysis reveals that the broadening of the effective transmission induced by environmental dephasing also extends into the energy gaps separating the bulk bands. The corresponding transmission amplitudes are small, become discernible only on a logarithmic scale (not shown here). This indicates the presence of weak incoherent tunneling through otherwise forbidden energy regions. \vspace{0.15cm}

\section{Closing Remarks}
\label{sec:remark}
In summary, we have presented a comprehensive analysis of the spectral and transport properties of the commensurate off-diagonal Aubry-André-Harper (AAH) chain, focusing on the representative case of $b=1/6$ that exemplifies the topologically nontrivial, gapless regime. Through combined spectral and transport investigations, we identified the first topological transition at $W_{c1}=\frac{2}{\sqrt{3}}$, where the zero-energy edge state disappears due to the complete overlap of the central bands. A second transition at $W_{c2}=2$ marks the gap closing between a central and adjacent bulk band, accompanied by the hybridization of quantum Hall edge states with the bulk continuum.

Even-odd lattice parity plays a decisive role in transport: chains with $2qn+1$ exhibit robust, perfectly ballistic conduction at all Dirac points including the zero-energy state at first transition point. Besides this, configurations with $N=6n+5$ display high transmission at $E=\pm t_0$ at the second transition. These features persist across larger `q' values, though the exact transition energies and optimal lattice sizes shift accordingly.

Furthermore, introducing environmental dephasing through Büttiker probes leads to spectral broadening and a marked enhancement of transmission near the band center, particularly for weak probe coupling ($\tau_p \le 0.5$). This demonstrates that moderate decoherence can facilitate incoherent yet efficient charge transport, preserving high-conducting behavior even beyond the topological transition points.

These results provide direct insights into how modulation parameters and decoherence jointly dictate electron dynamics in modulated lattices. The predicted transitions and transport signatures can be experimentally probed in photonic waveguide arrays, cold-atom optical lattices, or semiconductor superlattices, where tunable hopping and phase control are routinely achievable. 
Our findings thus offer a realistic pathway to observing topological phase transitions and coherence-decoherence crossover in engineered one-dimensional quasiperiodic systems.

\section*{ACKNOWLEDGMENTS}

The author would like to thank M. Saha, J. Majhi, and S. Basak for fruitful discussions. This work was supported by the TRC project (RA-I) funded by the Department of Science and Technology, Government of India, at the School of Physical Sciences, Indian Association for the Cultivation of Science.

\bibliographystyle{apsrev4-2}
\bibliography{reference}

\end{document}